%% file: main.tex
\documentclass[journal]{IEEEtran}
\usepackage[english]{babel}
\usepackage{adjustbox}
\usepackage[caption=false,font=normalsize,labelfont=sf,textfont=sf]{subfig}
\usepackage[draft]{hyperref}
\usepackage{alphalph}
\usepackage{xcolor}
\usepackage{tabularx}
\usepackage{colortbl} 
\usepackage{graphicx}
\def\BibTeX{{\rm B\kern-.05em{\sc i\kern-.025em b}\kern-.08em
    T\kern-.1667em\lower.7ex\hbox{E}\kern-.125emX}}

\usepackage{comment}

\usepackage{booktabs}

\providecommand{\keywords}[1]
{
  \small	
  \textbf{\textit{Keywords---}} #1
}

\begin{document}

\title{Internet of Things Security: A Survey on Common Attacks}
\author{Dalton Cézane Gomes Valadares (IEEE Senior Member), Luiz Antonio Pereira Silva, Daniel Hindemburg de Miranda Marques, Álvaro Alvares de Carvalho César Sobrinho, Andson Marreiros Balieiro, Mohamed Ahmed Hail, Mohammed B. Alshawki, and Kyller Costa Gorgônio
\thanks{Dalton C. G. Valadares, \textit{Universidade Federal da Paraíba} (UFPB). E-mail: daltoncezane@gmail.com.}
\thanks{Luiz A. P. Silva, \textit{Universidade Federal de Campina Grande} (UFCG)}
\thanks{Daniel H. M. Marques, \textit{Agência Nacional de Telecomunicações} (Anatel) and \textit{Universidade Federal de Campina Grande} (UFCG)}
\thanks{Álvaro
A. de C. C. Sobrinho, \textit{Universidade Federal do Agreste de Pernambuco} (UFAPE)}
\thanks{Andson M. Balieiro, \textit{Universidade Federal de Pernambuco} (UFPE)}
\thanks{Mohamed A. Hail, \textit{University of Lübeck}}
\thanks{Mohammed B.
Alshawki, \textit{Eötvös Loránd University} (ELTE) and \textit{Hochschule Furtwangen}}
\thanks{Kyller C. Gorgonio, \textit{Universidade Federal de Campina Grande} (UFCG)}
}

\markboth{IEEE,~Vol.~XXX, No.~XXX, Month~202X}%
{IoT Security}

\maketitle

\begin{abstract}\label{Abstract}
The exponential growth of the Internet of Things (IoT) has integrated connected devices into various sectors like smart cities, digital health, and Industry 4.0, generating vast amounts of real-time data to support intelligent decision-making. However, this widespread adoption is fundamentally challenged by significant security risks, primarily due to the inherent computational limitations of devices, lack of standardization, and an expanding attack surface. Given that security is paramount to ensuring trust in these environments, this paper presents a comprehensive survey and a multi-dimensional analysis of the IoT threat landscape. It describes 28 common attacks, ranging from traditional threats, such as Man-in-the-Middle, to specialized IoT exploits, including node replication and skimming. To provide a structured understanding of these risks, we employ the STRIDE model for functional threat classification alongside the CVSS framework for quantitative criticality assessment. Furthermore, the research establishes a robust mapping between these threats and five foundational vulnerability classes (Process, Code, Communication, Operation, and Device), uncovering the specific technical entry points exploited by adversaries. Beyond threat identification, the survey presents state-of-the-art mitigation techniques and discusses emerging paradigms and research gaps, working as a roadmap for future investigation and providing a consolidated technical foundation for both researchers and practitioners aiming to build resilient and secure IoT ecosystems.    
\end{abstract}

\keywords{Internet of Things, Security, Attacks, Vulnerabilities, IoT}

\input{Introduction}

\input{Background}

\input{RelatedWork}

\input{threatmodel}

\input{AttackDetail}

\input{StrideCVSS}

\input{vulnattacks}

\input{vertattacks}

\input{FutureDirections}

\input{Conclusion}

\section*{Acknowledgment}

The authors thank \textit{Fundação Coordenação de Aperfeiçoamento de Pessoal de Nível Superior (Capes)} for partial financing of this work.
\vspace{-3mm}

\medskip

\bibliographystyle{IEEEtran}
\bibliography{ref.bib}

\end{document}

%% file: Introduction.tex
\section{Introduction}
\label{sec:Intro}
More than 20 years after the term Internet of Things (IoT) was defined~\cite{historyofinfo3411}, its use continues to grow in in different contexts, driven by the popularization of connected devices, advances in wireless networks, and the increasing demand for automation in different sectors. IoT allows physical objects, such as sensors, cameras, appliances, and vehicles, to communicate with each other and with systems, generating real-time data and enabling intelligent decisions. Today, this technology enables scenarios, such as smart cities, digital health, precision agriculture, industry 4.0, and home automation. The scale of this deployment is unprecedented, with recent estimates suggesting there will be approximately 30 billion connected IoT devices globally by 2030~\cite{sinhaIoTlytcs_25}, creating a ubiquitous and highly complex digital ecosystem.

According to a Grand View Research report~\cite{grandview2024iot}, the global IoT market was valued at approximately \$1.18 trillion in 2023 with a prediction of reaching approximately \$2.65 trillion in 2030, with a compound annual growth rate (CAGR) of 11.4\% during the period. McKinsey \cite{mckinsey2021iot} also points out that the IoT could generate between \$5.5 and \$12.6 trillion in global economic value by 2030, especially in sectors such as manufacturing, healthcare and retail. Investments in IoT also reflect this growth. IDC (International Data Corporation)~\cite{idc2023iot} estimates that global spending on IoT should exceed \$1 trillion in 2026, driven by areas such as manufacturing operations and production asset management. In addition, with the expansion of 5G networks and edge computing, the volume of data processed locally is expected to improve significantly, further expanding the potential of IoT in critical applications, such as autonomous vehicles and real-time medical monitoring.

Considering the current technological trends and the investment predictions across diverse IoT domain applications, data security is a primary challenge to the widespread adoption and trust. Due to the limited computational capability of IoT devices~\cite{valadares21_slr}, the implementation of robust encryption, authentication, and secure firmware update mechanisms remains challenging. Many IoT devices are also improperly configured, often employing weak passwords or leaving unnecessary ports open, which significantly increases their exposure to cyber threats. Furthermore, device heterogeneity, coupled with the lack of standardization and inadequate secure development practices, can significantly expand the attack surface~\cite{valadares2023systematic}. This heterogeneous nature often leads to fragmented security implementations, where a single weak link can compromise entire networks. The possibility of different settings has further exacerbated these risks, as legacy systems with minimal security are now exposed to internet-based threats.
  
As a result, IoT devices are frequently exploited by attackers~\cite{valadares21_slr,valadares23_aina}. One notable example is the Mirai botnet~\cite{antonakakis_mirai_17}, which compromised thousands of connected devices to perform massive DDoS attacks in 2016. These vulnerabilities are largely driven by weak authentication mechanisms and unnecessary Internet exposure. Since Mirai, IoT botnets have evolved significantly, becoming more sophisticated by exploiting Zero-Day vulnerabilities and employing command-and-control obfuscation techniques~\cite{affinito2023,Asadi24}. Modern threats have also shifted towards critical infrastructures. For instance, ransomware attacks targeting IIoT gateways can halt production lines, causing severe economic and physical impact~\cite{AlHawawreh19}. Furthermore, the IoT supply chain has emerged as a critical vector, where compromises in third-party software components or hardware manufacturing can introduce untrusted vulnerabilities before devices are even deployed~\cite{zhao22}. In addition, inefficient or nonexistent firmware update mechanisms~\cite{choudhary21} prevent timely correction of known vulnerabilities, perpetuating security risks throughout the device lifecycle.

Given the growing concerns about security in IoT environments, we conducted a literature survey to identify common attacks targeting IoT devices and applications. Despite the existence of several surveys addressing IoT security, the majority tend to focus either on specific architectural layers or on isolated mitigation paradigms, such as Machine Learning or Blockchain. There is a critical lack of a holistic, multi-dimensional taxonomy that bridges the gap between theoretical vulnerabilities and practical impact assessments. To build resilient systems, stakeholders require a unified framework that correlates a comprehensive set of attacks with their functional impacts, severity levels, and precise architectural entry points. 

Thus, this survey covers recent scientific literature, including papers from recognized journals, international conferences, and technical reports indexed in major scientific databases (such as IEEE Xplore\footnote{https://ieeexplore.ieee.org/}, ACM Digital Library\footnote{https://dl.acm.org/}, and Springer Link\footnote{https://link.springer.com/}), focusing on identifying and categorizing the attacks most frequently reported in IoT security studies. We compiled a set of 28 common attacks, ranging from classic threats such as man-in-the-middle and denial-of-service (DoS) to IoT-specific threats, such as skimming and node replication attacks. Based on this identification, we also sought to map techniques and best practices proposed in the literature to mitigate these attacks. The strategies range from lightweight authentication mechanisms and encryption adapted for low-power devices to the use of machine learning to detect anomalous behavior, in addition to security recommendations at the network, firmware, and application levels. This work thus offers a useful basis for both researchers and practitioners in IoT Systems by providing a consolidated view of the predominant threats and the most promising technological responses.

To provide a systematic evaluation of the identified threats, this work classifies the 28 common IoT attacks using the STRIDE model and the Common Vulnerability Scoring System (CVSS). The STRIDE methodology is employed to categorize threats based on their functional impact, specifically addressing Spoofing, Tampering, Repudiation, Information Disclosure, Denial of Service, and Elevation of Privilege. This classification is further enriched by the application of CVSS scores, which quantify the severity of each attack, ranging from Medium and High to Critical levels. By combining these two frameworks, the paper offers a robust technical foundation that not only identifies the nature of the vulnerabilities but also prioritizes them based on their potential impact on the IoT ecosystem.

Furthermore, recognizing that the threat landscape is continuously evolving, this survey goes beyond current countermeasures, briefly discussing emerging security paradigms. For instance, we explore the transition towards Zero Trust architectures, AI-assisted anomaly detection, and the architectural shift towards IoT over Named Data Networking (IoT-NDN) as a promising data-centric security solution. This study also incorporates a structured vulnerability classification scheme that considers the following five classes based on their origin and nature: Process, Code, Communication, Operation, and Device. Considering this classification, we mapped each of the 28 described attacks to these specific vulnerability classes to identify the underlying security weaknesses and technical entry points exploited by malicious actors. This mapping provides a coordinated defense framework, allowing for the strategic alignment of mitigation responsibilities, ranging from hardware-level fixes to operational security monitoring, thereby contributing to the development of a more resilient IoT environment.

\subsection{Contributions}
\label{sec:contr}

\begin{itemize}
    \item \textbf{Comprehensive identification of threats} - The study provides a compiled set of 28 common attacks relevant to IoT environments, ranging from classic network threats like Man-in-the-Middle and Denial-of-Service to IoT-specific attacks such as Skimming and Node Replication.

    \item \textbf{Threat classification and severity assessment} - Each identified attack is rigorously categorized using the STRIDE model, mapping them to functional threat categories (Spoofing, Tampering, Repudiation, Information Disclosure, Denial of Service, and Elevation of Privilege), and evaluated with the Common Vulnerability Scoring System (CVSS), providing a criticality level enabling prioritization based on the scores.

    \item \textbf{Attack-Vulnerability mapping} - The research establishes a direct correlation between the 28 attacks and five distinct vulnerability classes (Process, Code, Communication, Operation, and Device), which allows for a better visualization of the technical entry points exploited by attackers.

    \item \textbf{Mitigation strategies} - For every described attack, the paper identifies specific techniques and best practices proposed in recent literature to mitigate these threats, including lightweight encryption, machine learning for anomaly detection, and secure routing protocols.

    \item \textbf{Future directions} - The survey presents technologies such as Named Data Networking (NDN) as a promising paradigm shift for next-generation IoT security, highlighting its data-centric security advantages over traditional IP-based architectures.

    \item \textbf{Identification of research gaps} - The work outlines current challenges such as resource constraints, device lifecycle security, and the lack of standardized monitoring tools, offering a roadmap for future research in the field.
\end{itemize}

\subsection{Survey structure}
\label{sec:struc}

The remainder of this paper is organized as follows: Section \ref{sec:back} establishes the foundational background, detailing the layered IoT architecture, primary application verticals, core security concepts, and the STRIDE and CVSS frameworks; Section \ref{sec:RelatedWork} reviews related work, highlighting the current state-of-the-art and the gaps addressed by our comprehensive taxonomy; Section \ref{sec:threat} defines the threat model and analyzes the attack surfaces across the Perception, Network, and Application layers; Section \ref{sec:AttackDetail} presents an in-depth analysis of 28 common IoT attacks, along with their respective mitigation strategies; Section \ref{sec:stridecvss} classifies these attacks according to the STRIDE categories and evaluates their severity using the CVSS scoring system; Section \ref{sec:vulnattacks} introduces a novel taxonomy that maps the identified attacks to five core vulnerability classes and specific architectural entry points, providing guidelines for proactive security planning and reactive monitoring; Section \ref{sec:vert} contextualizes the threat landscape by mapping the most prevalent attacks to four major IoT verticals; Section \ref{sec:FutureDirections} discusses ongoing challenges, research gaps, and emerging opportunities, emphasizing the role of Artificial Intelligence and the transition towards IoT over Named Data Networking (IoT-NDN); finally, Section \ref{sec:Conclusion} concludes the survey and outlines critical directions for future research.

%% file: Background.tex
\section{Background}
\label{sec:back}
This section establishes the theoretical and architectural foundations necessary to contextualize our study. It delineates a common layered IoT architecture, identifies key application verticals, and defines core security principles. Furthermore, it introduces the STRIDE threat model and the Common Vulnerability Scoring System (CVSS), which serve as the qualitative and quantitative methodological pillars for our subsequent analysis.

\subsection{Common Architecture for IoT Applications}
\begin{figure*}[!thb]
    \centering
    \includegraphics[width=0.7\textwidth]{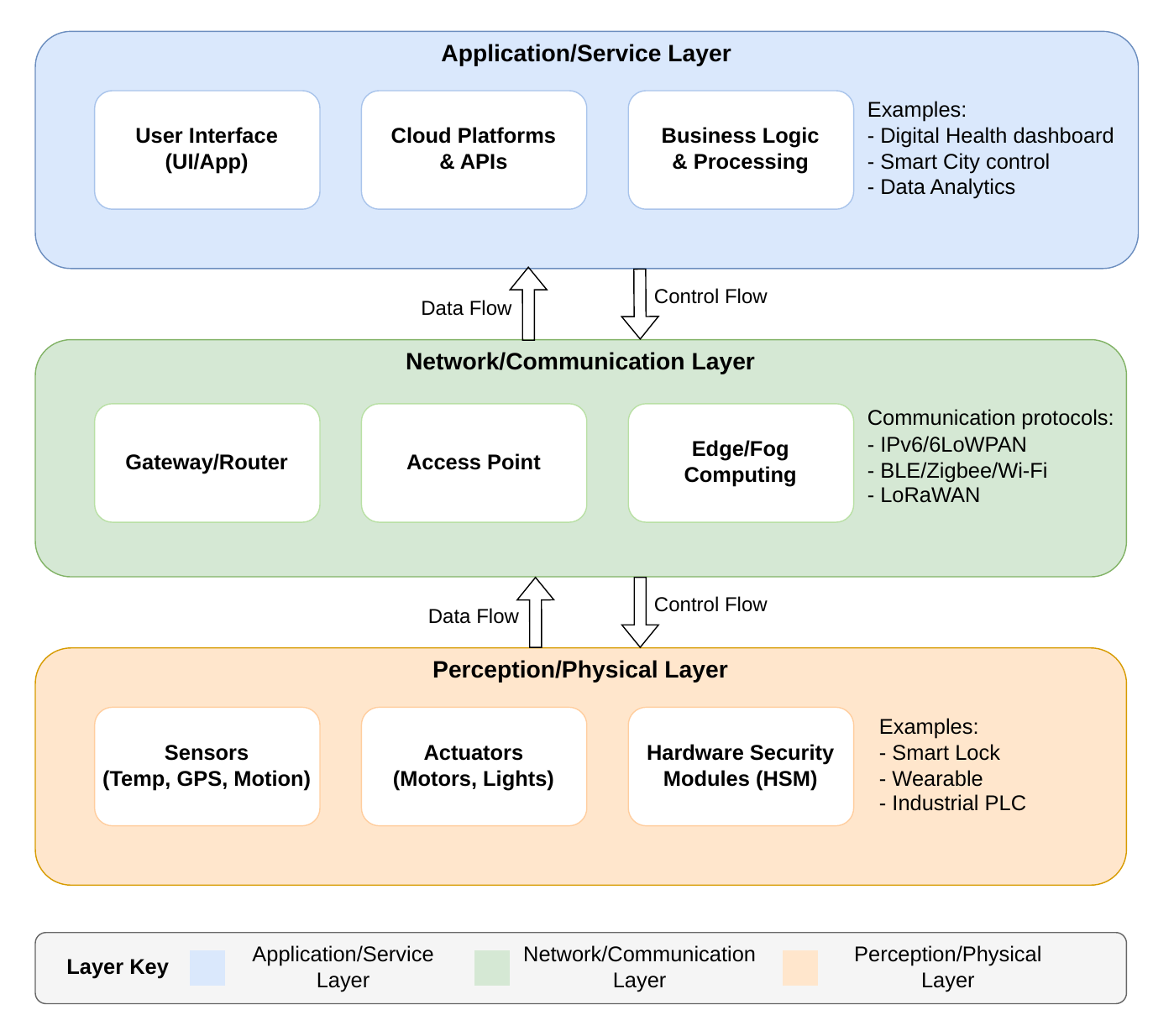}
    \caption{Layered Architecture for IoT Applications.}
    \label{fig:arch}
\end{figure*}

To provide a comprehensive understanding of the IoT security landscape and contextualize the attacks discussed in this paper, it is essential to outline the common architectural layers that form the foundation of most IoT applications. As depicted in Figure \ref{fig:arch}, IoT systems are typically structured into a tiered model, moving from data acquisition in the physical world to user services in the digital world. This multi-layered structure is critical for managing the complexity, heterogeneity, and scale of modern IoT deployments.

At the lowest tier, the Perception Layer, also referred to as the Physical or Sensing Layer, consists of the varied and numerous ``things'': the hardware devices. As shown in the figure, this layer considers a multitude of nodes, sensors, and actuators responsible for directly interacting with and collecting data from the environment. This foundational tier performs the core functions of data acquisition and initial processing, transforming physical signals into digital information. These devices are often resource-constrained, prioritizing energy efficiency and low cost, which significantly impacts their capability for robust on-device security measures.

Positioned above the perception layer is the Network Layer, also known as the Gateway or Communication Layer. This intermediate tier, illustrated by gateways and routers in the figure, acts as the critical bridge connecting the physical devices to the digital infrastructure. The network layer's primary responsibility is data transmission and routing. It facilitates the flow of aggregated data from various sensory nodes at the perception layer across diverse communication networks (e.g., Wi-Fi, 6LoWPAN, Zigbee, LTE, LoRaWAN) to the core processing systems. Furthermore, gateways within this layer often perform protocol translation, standardizing communication between heterogeneous IoT devices and the backend.

The topmost tier is the Application Layer, where user services, data analytics, and cloud platforms reside. This layer ingests the data collected and transmitted from the layers below. It performs high-level functions such as data processing, storage, and complex analytics to generate valuable insights and automate actions. The application layer provides user-facing interfaces, dashboards, control mechanisms, and APIs (Application Programming Interfaces) for system management and interaction. It is also where critical security functions, such as authentication and access control, are typically centralized for robust access management to cloud resources and services. This separation allows for a coordinated defense framework, addressing security responsibilities from the physical devices up to user-facing applications and data governance within the broader IoT ecosystem.

\subsection{Common Verticals for IoT Applications}

The versatility of IoT technology allows its integration into numerous sectors, shifting paradigms from manual operations to data-driven and intelligent automation. Based on current market trends and economic projections, four primary verticals stand out as the most significant drivers of IoT adoption, which are described as follows.

\subsubsection{Smart Cities}
This vertical focuses on the demand for automation in urban environments to improve the quality of life and operational efficiency. IoT applications in this sector include intelligent traffic control, waste management, public safety through connected cameras, and energy-efficient smart lighting.

\subsubsection{Digital Health (Healthcare)} IoT has revolutionized medical services through real-time monitoring and digital health applications. Also referred to as the Internet of Medical Things (IoMT), this vertical utilizes wearable devices to collect personal health data and connected medical systems to enable intelligent clinical decisions. The integration of 5G and edge computing is particularly critical here, enabling high-speed data processing for real-time medical monitoring.

\subsubsection{Industry 4.0 (Industrial IoT)} Often referred to as the Industrial Internet of Things (IIoT), this vertical is one of the largest recipients of global spending. It focuses on manufacturing operations, production asset management, and supply chain optimization. This sector serves as a primary engine for global industrial transformation, representing a cornerstone of the modern global economy, with the potential to generate significant financial returns through the synergy of self-optimizing systems and proactive, predictive maintenance frameworks.

\subsubsection{Precision Agriculture (Agriculture 4.0)} Driven by the growing demand for food security and automation in the primary sector, IoT is used to monitor soil conditions, weather patterns, and livestock health. These applications generate real-time data that allow farmers to make intelligent decisions regarding irrigation, fertilization, and harvesting, maximizing yield while minimizing resource waste.

\subsection{Security Concepts}
To establish a comprehensive understanding of the threats impacting IoT environments, it is essential to define the fundamental security concepts that serve as the pillars of information protection. These concepts, often referred to through the CIA triad (Confidentiality, Integrity, and Availability)~\cite{sobrinho24}, are used to evaluate the impact of the 28 attacks identified in this study. Their definitions are presented as follows.

\subsubsection{Confidentiality} This principle ensures that sensitive information is accessible only to authorized entities and is protected from unauthorized disclosure. In the context of IoT, this involves safeguarding personal data collected by wearables or operational information from industrial sensors against attacks such as eavesdropping.

\subsubsection{Integrity} This refers to the protection of data against unauthorized or accidental modification, ensuring that the information remains accurate and reliable throughout its lifecycle. Attacks like tampering or node replication directly target this principle by altering device codes or transmitted parameters.

\subsubsection{Availability} This ensures that systems, networks, and data are consistently accessible and usable by authorized users when needed. In IoT, availability is a critical challenge due to resource-constrained devices being susceptible to Denial of Service (DoS) and battery exhaustion attacks.

\subsubsection{Authenticity} This involves verifying the identity of a user, device, or system participating in a communication. Weak authentication mechanisms are a primary vulnerability in IoT, allowing for spoofing and masquerade attacks where an adversary impersonates a legitimate node.

\subsubsection{Accountability and Non-Repudiation} Accountability ensures that every action within a system can be traced back to a specific actor, while non-repudiation prevents an entity from denying an action it has performed. These are vital for auditing and digital forensics in distributed IoT networks.

\subsection{STRIDE: Threat Classification Model}

The STRIDE model~\cite{valadares23_aina} is a methodology developed by Microsoft to identify and classify security threats in computer systems. It is an acronym formed by the initials of six threat categories. Spoofing, Tampering, Repudiation, Information Disclosure, Denial of Service, and Elevation of Privilege. The model has been widely used in the threat analysis phase of secure software engineering processes, including Internet of Things (IoT) contexts, where the diversity of devices and the complexity of interactions expand the attack vectors.

Each STRIDE category is associated with a specific type of flaw or malicious behavior:
\begin{itemize}
    \item Spoofing - Refers to identity falsification, where an attacker impersonates another user, device, or system. An example in IoT would be a fake sensor registering on a network without proper authentication.
    \item Tampering - This involves the malicious modification of data in transit or at rest. In IoT environments, this can occur, for example, when changing commands sent to an industrial actuator.
    \item Repudiation - This involves actions that cannot be attributed to a specific actor, due to the lack of reliable logs or records. This problem makes accountability and auditing difficult, especially in distributed systems.
    \item Information disclosure - This refers to the leaking of sensitive information. In IoT, this can include personal data collected by wearable devices or operational information from an industrial plant.
    \item Denial of Service (DoS) - Attacks that aim to disrupt the operation of a service or system. IoT devices with few resources are particularly susceptible to denial of service attacks.
    \item Elevation of Privilege - When attackers gain more privileges than they should, allowing them to perform actions that are normally restricted. In IoT, this can mean taking full control of a device from a firmware exploit.
\end{itemize}

Using STRIDE in threat modeling processes provides a systematic approach to identifying vulnerabilities in each component of the system, based on its role and associated data flows. By applying the model, security teams can prioritize risks, propose mitigation controls, and document security assumptions, contributing to the construction of more resilient architectures.

In IoT environments, STRIDE is especially useful because it provides a framework that can be adapted to multiple layers of the architecture, from the device to the cloud, enabling the detection of threats in communication protocols, software interfaces, firmware, and even authentication processes.

\subsection{Common Vulnerability Score System (CVSS)}

The Common Vulnerability Scoring System (CVSS)~\cite{cvss40} provides a numerical representation for the severity of a security vulnerability. It is a vendor-neutral, industry-standard framework designed to convey the principal characteristics of a vulnerability and produce a score reflecting its technical risk. By using a consistent scoring methodology, organizations can prioritize their vulnerability management processes and ensure that the most critical threats are addressed with urgency.

The CVSS 4.0 considers the following four distinct metric groups:
\begin{itemize}
    \item Base - Represents the intrinsic qualities of a vulnerability that are constant over time and across different user environments;
    \item Threat - Reflects the characteristics of a vulnerability that change over time, such as the availability of exploit code or analytical reports;
    \item Environmental - Represents characteristics that are unique to a specific user's environment, allowing for the customization of the score based on the importance of the affected asset;
    \item Supplemental - Provides additional insights into the characteristics of a vulnerability without modifying the final numerical score.
\end{itemize}

The assessment process begins with the Base metric group, which is evaluated by analysts to produce a score ranging from 0.0 to 10.0. This group is composed of two primary sets: Exploitability metrics, which reflect the ease with which a "vulnerable system" can be compromised, and Impact metrics, which measure the direct consequences to both the vulnerable system and any ``subsequent systems'', including potential impacts on human safety. While Base metrics are typically defined by vendors or bulletin analysts with detailed technical knowledge of the vulnerability, they assume a worst-case scenario for threat and environmental factors by default.

To achieve a more precise reflection of risk, the initial assessment can be refined through the Threat and Environmental metric groups. The Threat group accounts for factors that change over time, such as the public availability of proof-of-concept exploit code, which can lower the final score if no active threat is confirmed. Simultaneously, the Environmental group allows consumer organizations to customize the score based on their specific infrastructure, considering existing security controls and the relative importance of the affected system. While assessing these groups is not mandatory, it is highly recommended to produce meaningful results tailored to a specific point in time and a particular computing environment.

Finally, the Supplemental metric group provides additional context without directly modifying the final numerical score. These metrics allow consumers to apply their own locally significant severity levels to attributes that do not fit within the traditional scoring groups. Alongside the quantitative score, the assessment produces a vector string, a specifically formatted textual representation of all assigned values. This vector string is essential for transparency, as it conveys the exact characteristics used to derive the vulnerability's qualitative rating and should always be displayed with the final score.

The CVSS defines five severity levels for a vulnerability: None (score 0.0); Low (score between 0.1 and 3.9); Medium (score between 4.0 and 6.9); High (score between 7.0 and 8.9); and Critical (score between 9.0 and 10.0). The classification and criticality score assist in the risk management and prioritization process resulting from threats and vulnerabilities, helping to mitigate them.

%% file: RelatedWork.tex
\begin{table*}[!th]
\centering
\caption{Comparison of our survey with existing related work in IoT Security}
\label{tab:related_work_comparison}
\renewcommand{\arraystretch}{1.2}
\small
\begin{tabularx}{\textwidth}{|l|X|c|c|c|X|}
\hline
\textbf{Reference} & \textbf{Main Focus} & \textbf{STRIDE/CVSS} & \textbf{Layer/Class Mapping} & \textbf{Mitigations} & \textbf{Key Technologies} \\ \hline

Butun \textit{et al.} \cite{butun2019security} & Attacks against WSN in IoT and countermeasures & No & No & Yes & Traditional WSN Security \\ \hline

Mrabet \textit{et al.} \cite{mrabet2020survey} & OSI model-based study of IoT security & No & Yes (Layered) & No & Cloud and Data Services \\ \hline

Malhotra \textit{et al.}  \cite{malhotra2021internet} & Utilization of ML to detect attacks in IoT systems & No & No & Yes & Machine Learning (ML) \\ \hline

Abdullahi \textit{et al.}  \cite{abdullahi2022detecting}  & Systematic review of AI/DL for threat detection & No & No & No & Artificial Intelligence (AI) \\ \hline

Khan \textit{et al.} \cite{khan2022internet} & IoT security with Blockchain technology & No & No & Yes & Blockchain \\ \hline

Blinowski \textit{et al.} \cite{blinowski2020cve} & CVE-based classification of vulnerable IoT systems & No & Partial & No & Machine Learning (ML) \\ \hline

\textbf{Our Survey} & \textbf{Multi-dimensional analysis of 28 IoT attacks} & \textbf{Yes (Both)} & \textbf{Yes} & \textbf{Yes} & \textbf{AI/ML, IoT-NDN, PQC, Zero Trust} \\ \hline

\end{tabularx}
\end{table*}

\section{Related work}
\label{sec:RelatedWork}

The study of attacks to IoT systems and the relevant mitigation is one of the most vital field that is continuously growing. This is due to the fact that the applications are also significantly increasing. Butun et al. \cite{butun2019security} studied 
the known attacks against WSN in IoT and their countermeasures. Later, Mrabet et al. \cite{mrabet2020survey} introduced an extended work by providing an OSI model based study of the IoT security, and also including an extra cloud and data layer. Their work included the study of the security of devices and sensors, network and communication layers, the application layer, and finally the cloud based data services. 
The previous results have been further supported by Malhotra et al. \cite{malhotra2021internet} which explained how ML can be utilized to detect attacks in IoT systems. Their result also compared different approaches that can be used to detect the attacks and discussed the strengths and weaknesses of those approaches.
Oluwalola and Oluyemi \cite{oluwalola2025introduction} continued the Butun's study by focusing on the existing regulations and the mechanisms such as cryptographic algorithms to protect the IoT network.

The recent evolution of IoT security research has increasingly involved the Artificial Intelligence (AI) and Deep Learning (DL) techniques \cite{sharma2023anomaly} to address the growing sophistication of attacks. Abdullahi et al. \cite{abdullahi2022detecting} through their systematic literature review provided the approaches to detect the attacks in IoT networks and also highlighted the general adoption of AI threat detection based on the categories of the attacks. Beyond traditional Machine Learning (ML) approaches \cite{rafique2024machine}, the research results currently employ deep neural networks, convolutional neural networks (CNNs), recurrent neural networks (RNNs), and long short-term memory (LSTM) architectures to detect zero-day attacks \cite{ibrahim2023anomaly} and anomalous behaviors in highly dynamic IoT environments. 
Federated Learning (FL) \cite{mothukuri2021federated} has also emerged as a promising paradigm, allowing distributed IoT devices to collaboratively train detection models without sharing raw data \cite{yan2024fedlabx}, which result in a preserving privacy approach, and at the same time maintain the detection accuracy. Furthermore, reinforcement learning techniques have been introduced to enable adaptive intrusion detection systems (IDS) \cite{hsu2020deep} that dynamically adjust their defense strategies based on evolving threat landscapes. These approaches provides a higher performance result in large-scale and heterogeneous IoT deployments \cite{zuech2015intrusion}. 

As highlighted by Khan et al. \cite{khan2022internet} in their state of the art review, in addition to AI-driven approaches, blockchain technology has also been utilized as a main building block to provide a decentralized security mechanism for IoT systems. Blockchain-based frameworks aim to prevent possible attacks through device authentication \cite{almadani2023blockchain} and data integrity. Additionally, Liu et al. \cite{liu2023survey} provided a survey on the utilization of blockchain technology to provide trust management without relying on centralized authorities, and in some cases in a privacy preserving approach \cite{valadares2023privacy}. Pervez et al. \cite{pervez2018comparative} performed a comparative analysis of blockchain protocols and Directed Acyclic Graph (DAG)-based distributed ledgers that have been proposed to overcome the computational and energy constraints of IoT devices. Smart contracts are utilized to automate access control policies and enforce security rules transparently \cite{isazade2025blockchain}. Recent research also explores hybrid architectures combining blockchain with edge computing \cite{douiba2026distributed}, where security operations are partially offloaded to edge nodes to reduce latency and resource consumption. 

Another fundamental approach in studying the IoT attacks are through CVE based analysis \cite{blinowski2020cve}. The standardized vulnerability datasets are utilized to identify common security weaknesses, analyze attack trends, and develop new techniques for vulnerability detection and mitigation. Blinowski et al. \cite{blinowski2020cve} provided a data-driven classification of vulnerabilities that affect the IoT network. This work demonstrated how ML can be utilized to classify vulnerabilities using CVE datasets. Kim and Yoo \cite{kim2022analysis} provided a statistical analysis of IoT vulnerabilities using CVE datasets. Their work provided a systematic mapping of CVE vulnerabilities and CWE categories of IoT devices, which also highlighted the dominant classes of vulnerabilities, such as memory corruption \cite{english2019exploiting}. Chen et al. \cite{chen2024empirical} linked the IoT devices with CVE vulnerability data in their survey, and investigated how the vulnerabilities affect the IoT devices and their relation to CVE records.

While previous efforts have significantly contributed to specific domains of IoT security -- such as mitigating attacks in WSNs, applying Machine Learning for anomaly detection, or leveraging Blockchain for decentralized trust -- most of these studies lack a holistic approach. As summarized in Table \ref{tab:related_work_comparison}, existing surveys typically do not combine functional threat models with quantitative severity assessments, nor do they map attacks to specific architectural entry points. To address this critical gap, our survey provides a comprehensive evaluation of 28 attacks using both the STRIDE and CVSS frameworks. Furthermore, we establish a novel taxonomy that correlates these threats with five vulnerability classes across three architectural layers, offering a structured foundation for deploying emerging paradigms like Zero Trust and IoT-NDN.

%% file: threatmodel.tex
\begin{figure*}[!tb]
    \centering
    \includegraphics[width=0.75\textwidth]{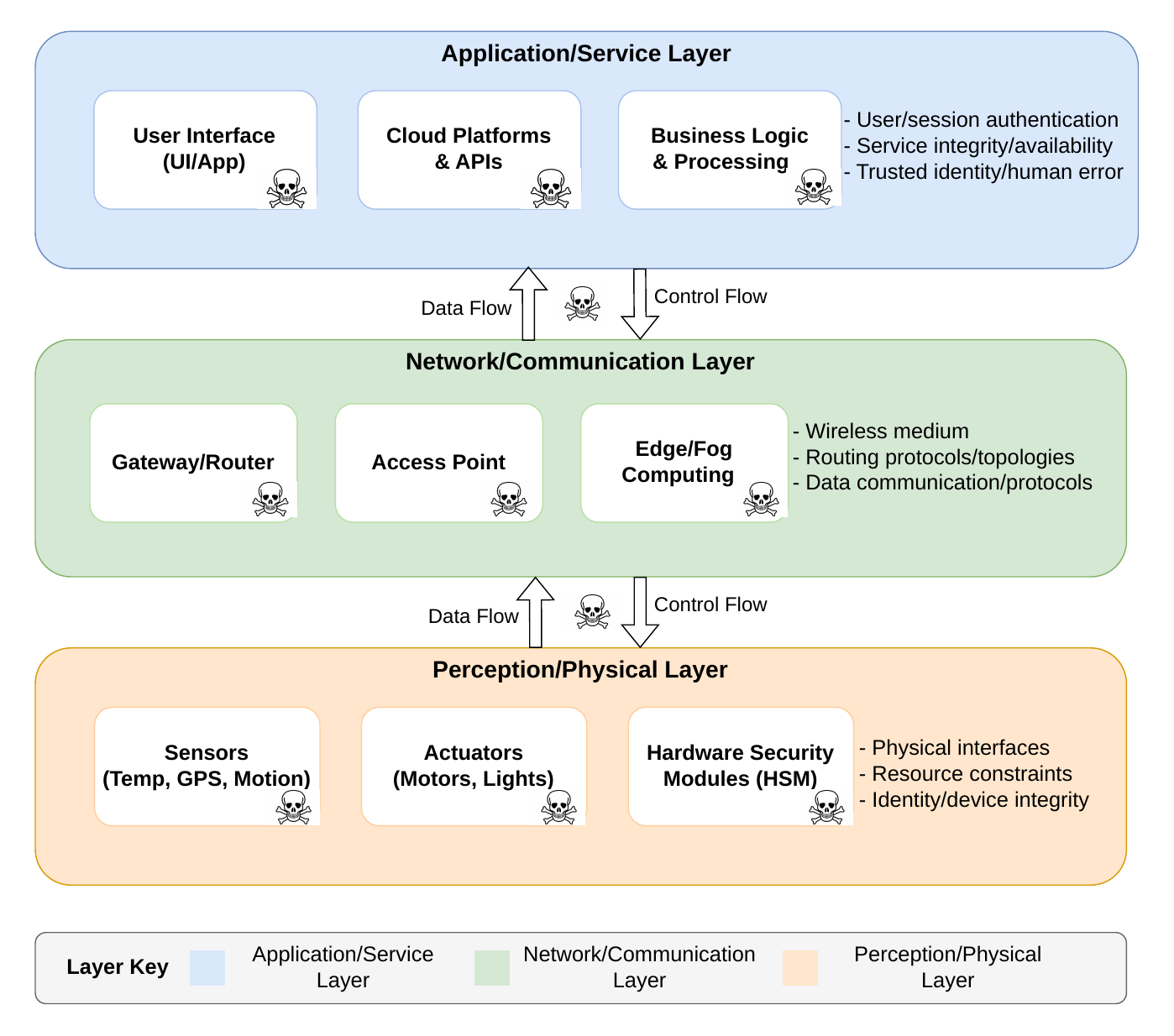}
    \caption{Threat Model.}
    \label{fig:threat}
\end{figure*}

\section{Threat Model and Attack Surfaces}
\label{sec:threat}

To design effective security mechanisms for IoT ecosystems, it is imperative to consider a comprehensive threat model that identifies potential adversaries, their capabilities, and the specific entry points they exploit. By leveraging the layered taxonomy presented in Fig. \ref{fig:arch}, this section analyzes the comprehensive attack surface of IoT applications, discussing how vulnerabilities in the Perception, Network, and Application layers translate into tangible security risks. The proposed model considers a hybrid adversary capable of performing both remote/logical attacks and local/physical manipulations. Fig. \ref{fig:threat} exhibits the threat model using skulls to represent the attack surfaces (e.g., devices and communication channels).

\subsection{Adversary Model and Capabilities}

We consider an adversary model tailored to the heterogeneous nature of IoT deployments. Adversaries can range from script kiddies utilizing automated tools to sophisticated, state-sponsored actors targeting critical infrastructure. Based on their operational context, we categorize adversaries into the following two primary types.

\begin{itemize}
    \item Remote adversaries - These actors lack physical access to the IoT deployment but can exploit logical vulnerabilities in the Network and Application layers. Their capabilities include intercepting wireless traffic, injecting malicious packets, launching volumetric DDoS attacks, and exploiting software flaws in cloud platforms or user interfaces.
    \item Local/physical adversaries - These actors possess direct physical access to IoT nodes or operate within the range of their wireless communication channels. Their capabilities include dismantling hardware to extract cryptographic keys, deploying malicious clones, performing radio frequency jamming, and installing physical skimming devices.
\end{itemize}
     
\subsection{Attack Surface Analysis by Layer}

The attack surface represents all points where an unauthorized user can attempt to enter or extract data from an environment. Our threat model decomposes this surface according to each of the three architectural layers: perception, network, and application.

\vspace{1em}
\subsubsection{Perception/Physical Layer}

This foundational layer, comprising hardware sensors, actuators, and edge devices, constitutes the most challenging attack surface due to the physical exposure of nodes and their severe resource constraints. The threat model considers the following three critical entry points in this layer.

\begin{itemize}
    \item \textbf{Physical hardware interface} - Unlike traditional IT assets secured within data centers, many IoT devices are deployed in public or unmonitored environments (e.g., smart streetlights, agricultural sensors). This exposure permits physical adversaries to tamper directly with the hardware. Thus, attackers may capture devices to extract sensitive data from flash memory or perform Node Replication to introduce malicious clones into the network. Furthermore, Skimming attacks target the physical interface of payment or access control devices to steal credentials directly.
    \item \textbf{Resource constraints (e.g., power and memory)} - adversaries exploit the limited computational and energy budgets of IoT nodes. By engineering interactions that keep nodes constantly active, attackers can launch Sleep Deprivation attacks to rapidly drain batteries, effectively destroying the node functionality without breaking any encryption. Similarly, Resource Exhaustion attacks flood constrained nodes with nonsense data to overwhelm their limited CPU or RAM, causing service failure at the edge.
    \item \textbf{Identity and device integrity} - The identity management and code integrity of local nodes represent a primary logical entry point. Adversaries can utilize Backdoors injected during manufacturing or via insecure update mechanisms to gain persistent, covert access. Additionally, Sybil attacks can exploit weak neighbor discovery processes, allowing a single malicious node to assume multiple false identities to undermine local routing or voting protocols.
\end{itemize}

\vspace{1em}
\subsubsection{Network/Communication Layer Attack Surface}

This intermediate layer acts as the communication backbone, bridging local perception devices with backend systems. Its attack surface is exceptionally dense, encompassing routing topologies, wireless channels, and data flow mechanisms. Threat actors exploit this surface to isolate devices, hijack communication, or manipulate data in transit.

\begin{itemize}
    \item \textbf{Wireless communication medium} - Nearly all IoT communication relies on wireless channels (e.g., Wi-Fi, Zigbee, BLE, LoRaWAN), which are inherently open to airwaves. This visibility can enable local adversaries to launch Jamming attacks to block legitimate signals, causing a pervasive Denial of Service (DoS). Remote or local attackers can exploit Bluetooth-specific flaws via Blueborne to take over devices without user interaction, or utilize De-authentication attacks to forcibly disconnect nodes from their access points. Passive adversaries can perform Eavesdropping on unencrypted channels to capture sensitive data, forming the basis for subsequent sophisticated attacks.
    \item \textbf{Routing protocols and network topology} - IoT networks, particularly multi-hop wireless sensor networks (WSNs), rely on decentralized routing protocols like RPL (Routing Protocol for Low-power and Lossy Networks). Threat actors can exploit the logic of these protocols to manipulate the network topology. An adversary can execute a Sinkhole attack, advertising the ``best'' path to the root to attract all traffic through itself for modification or selective dropping (Grayhole/Blackhole attacks). Advanced adversaries can create an out-of-band, low-latency tunnel between two compromised nodes (e.g., a Wormhole attack) to trick distant nodes into thinking they are neighbors, thereby distorting the entire network's perspective of its topology.
    \item \textbf{Data in transit and protocol logic} - Attackers can target the integrity and authenticity of packets as they traverse the network. Through Man-In-The-Middle (MITM) positions, adversaries can actively intercept, read, and modify data, violating confidentiality and integrity simultaneously. By executing Replay Attacks, outdated control messages can be retransmitted to trick actuators into performing unauthorized actions. Attackers can also utilize Spoofing (IP or ARP) to masquerade as trusted nodes to manipulate application-level logic or bypass access controls. Sophisticated Desynchronization attacks can break the state synchronization between a client and an access point, terminating active sessions and potentially forcing insecure re-authentication attempts.
\end{itemize}

\subsection{Application/Service Layer Attack Surface}
This topmost layer represents the primary interface for user interaction and high-level data governance, involving cloud platforms, dashboards, and APIs. The attack surface here is characterized by complex business logic, session management, and human-centric factors, allowing attackers to compromise the system remotely at a systemic scale.

\begin{itemize}
    \item User and session authentication - Authentication is one of the primary gatekeepers for application-level resources. Adversaries can target weak password policies via automated Brute-Force attacks. Furthermore, attackers can utilize social engineering (Phishing) or credential harvesting malware to facilitate Identity Theft, allowing them to Masquerade as legitimate users and gain unauthorized access to data and control interfaces across the entire deployment. Legitimate sessions can also be seized via Hijacking attacks, granting control without needing the original credentials.
    \item High-level service integrity and availability - The concentrated infrastructure of cloud platforms and main gateways may present an attractive target for systemic disruption. Adversaries can orchestrate Distributed Denial of Service (DDoS) attacks, often leveraging botnets composed of compromised IoT devices from the perception layer (like Mirai), to overwhelm application servers, rendering dashboards and control mechanisms inaccessible to legitimate users.
    \item Trusted identity and human error - The attack surface includes threats originating from within the trusted boundary. For instance, Insider Attacks by authorized users with malicious intent (or who are accidentally compromised) can cause devastating damage, as they already possess the necessary access privileges. Adversaries can also utilize Masquerade or Identity Theft to bypass organisational defences by appearing as trusted internal actors.
\end{itemize}

%% file: AttackDetail.tex
\section{Attacks in Internet of Things Scenarios}
\label{sec:AttackDetail}

\renewcommand*{\thesubsectiondis}{\AlphAlph{\value{subsection}}}



\subsection{Eavesdropping}

The eavesdropping attack~\cite{10731741,10.1145/3580789,10.1155/2016/4313475} represents a significant threat to the confidentiality of the service, as its primary objective is to intercept information and take data furtively. This attack vector enables the attacker to clandestinely monitor networks and exploit security vulnerabilities and weak connections between IoT devices and servers/gateways. Eavesdropping is facilitated by the use of radio wave-receiving antennas, which intercept transmitted signals and provide unauthorized access to the captured data. 

Moreover, the eavesdropping attack can also be perpetrated by leveraging the Remote Procedure Call (RPC) functionality, enabling the perpetrator to assume full control over the targeted IoT device. This control gives the hacker the ability to initiate the attack remotely from a separate computer connected to the network. Consequently, the eavesdropping attack can be executed through the exploitation of RPCs, affording the attacker considerable flexibility and control. 

It is important to note that hackers typically employ eavesdropping attacks by intercepting communications that lack adequate protection measures. In such instances, the absence of robust security mechanisms and encryption makes the communication susceptible to interception and compromise, providing an opportunity for malicious actors to exploit vulnerabilities and gain unauthorized access to sensitive data~\cite{10.1016/j.comnet.2025.111545}.

\vspace{1em}
\textbf{Mitigations}

An approach to mitigating eavesdropping is to establish a secure transmission channel between authorized users. This can be achieved through the utilization of artificial noise~\cite{10.3390/computers14010006} and encoding techniques~\cite{10.1109/TNSE.2025.3565302} to enhance the security of the communication channel. Furthermore, the transmission channels can be subdivided into multiple sub-channels, ensuring that the exchanged communications are not solely dependent on a single channel. By employing such channel segmentation, even if an eavesdropper gains access to a portion of the segmented transmission, there is no visibility into the other data being transmitted through the other divisions. Consequently, any potential losses resulting from an eavesdropper's intrusion would be significantly reduced, limiting their ability to access comprehensive information.

Another effective mitigation is the application of cryptographic principles. For example, encrypted authentication ensures that snooping or eavesdropping entities are unable to capture and record user credentials. Using encryption during the authentication process, sensitive information exchanged between the user and the system remains protected and confidential. 

\subsection{Brute-Force Attack}

Most IoT devices often retain default or easily guessable passwords, making them vulnerable to attackers who can exploit weak credentials through brute-force attacks, often employing simple scripts. Brute-force attacks~\cite{7932855,s23135941} involve systematically guessing passwords using common word combinations to gain unauthorized access to IoT networks or exploring all the possible combinations for the password. Once an attacker successfully breaches the network through a brute-force attack, the situation becomes more complex to address. The infiltrator can utilize the compromised devices to establish a botnet network, thereby enabling the execution of more sophisticated and destructive Distributed Denial of Service (DDoS) attacks in the future. The brute-force attack serves as a precursor to exploiting a vulnerability that can potentially escalate into a larger chain of compromises. 

\vspace{1em}
\textbf{Mitigations}

To mitigate brute-force attacks, it is imperative to implement robust cryptographic measures and enforce strong password systems in the overall IoT ecosystem with its associated protocols~\cite{10190544}. Moreover, it is crucial to emphasize the importance of avoiding default passwords, as they facilitate easier detection by attackers and grant access to multiple devices that employ the same password. By incorporating improved cryptography and password practices, the security posture of IoT applications can be enhanced, reducing the risk of successful brute-force attacks and subsequent exploitation of vulnerabilities. 

The authentication must use strong passwords. A strong password should consist of multiple characters, including numbers, upper- and lowercase letters, and symbols. Avoid common names, numbers, and easily identifiable dates, as automated systems can easily decipher such patterns. Additionally, implementing robust authentication mechanisms such as two-factor authentication (2FA), multi-factor authentication (MFA), and zero-trust models is essential to thwart brute-force attacks~\cite{10.1038/s41598-025-01080-5}. 2FA and MFA require users to provide additional credentials or undergo additional verification steps beyond just a password, thus adding an additional layer of protection against brute-force attacks. Zero-trust models focus on verifying each access request, regardless of the source or user's location, minimizing the potential for unauthorized access even if the password is compromised. 

\subsection{Distributed Denial of Service (DDoS)}

Distributed Denial-of-Service (DDoS) attacks~\cite{salim20_ddos,neira_23_ddos} have a significant impact on the availability of targeted systems by overwhelming them with a barrage of unwanted Internet traffic. These attacks exploit numerous compromised machines and Internet-connected devices, including IoT devices (e.g. cameras and even light bulbs), smartphones, personal computers, and network servers. In the context of a DDoS attack, a massive communication network (for example, 5G mMTC~\cite{valadares23_aina}) can amplify its potency, as infected packets transmitted by multiple devices have the potential to infect all devices they traverse, leading to the formation of a botnet~\cite{thanh2021survey_botnets,rasool22_botnets}. This threat becomes even more pronounced in the IoT realm, where multiple compromised devices and branches can be leveraged simultaneously. 

The convergence of 5G and IoT exacerbates the impact of DDoS attacks, given the presence of numerous compromised devices and the rapid transmission of packets to multiple targets. The high-speed transmission capabilities of 5G technology expedite the propagation of these malicious packets across the network, enabling DDoS attacks to be executed at an even faster pace compared to other types of networks. Consequently, the 5G-IoT junction amplifies the potential for devastating DDoS attacks, requiring robust security measures to mitigate their impact and ensure the uninterrupted availability of systems and services~\cite{SINGH2024100543}.

\vspace{1em}
\textbf{Mitigations}

To effectively mitigate DDoS attacks, IoT systems must demonstrate resilience, adapt to network dynamics, and potential device failures. This resilience ensures that the service remains available and can quickly recover operation in the event of an attack, minimizing the extent of damage inflicted. Implementing a robust network monitoring service is crucial in detecting DDoS attacks. By analyzing system logs, it becomes possible to identify patterns of excessive request volumes that indicate a DDoS attack, which aims to flood the network with requests. Detection in time allows prompt action, mitigating the impact before significant damage occurs~\cite{SINGH2024100543}. 

Access control mechanisms play an essential role in maintaining the security, protection, and overall confidentiality, integrity, and availability of the system. By employing stringent access controls, unauthorized access and malicious activities can be thwarted, safeguarding critical assets. This procedure includes implementing proper authentication and authorization mechanisms, ensuring that only authorized entities get access to sensitive resources. Furthermore, data encryption and secure communication protocols further enhance the security posture of IoT devices and applications, safeguarding the confidentiality of data in transit and at rest. By incorporating resilience measures, robust network monitoring, and access control mechanisms, IoT applications can effectively mitigate the impact of DDoS attacks, ensuring continuous service availability~\cite{s20113078,886455}.

\subsection{Jamming}

The jamming attack~\cite{10.1145/3722041.3723096,Prasad2022} is one of the most prevalent and severe threats that can disrupt the normal operations of IoT networks. Malicious actors employ this attack to rapidly drain the batteries of IoT nodes, blocking data transmissions, and forcing IoT devices to repeat their transmissions repeatedly. Jamming attacks do not require extensive technical expertise, becoming a substantial challenge as they exploit the vulnerability of wireless communication technologies, making them susceptible to interference. Jamming attacks rely on overwhelming the target system's receiver by transmitting interference signals with higher power or interference levels.

Furthermore, detecting the presence of a jammer within the network can be challenging since jammers can disguise themselves as legitimate users, making it difficult to distinguish their malicious activities. Even if detected, mitigating the impact caused by a jamming attack is not straightforward as it can have a lasting effect on the network. Additionally, jamming attacks create opportunities for more complex attacks, such as Denial of Service (DoS) attacks, further exacerbating the threat landscape. Addressing the threat posed by jamming attacks is of utmost importance to ensure the reliability and security of IoT networks, especially in critical contexts where timely and accurate communication is essential~\cite{10175509}.

\vspace{1em}
\textbf{Mitigations}

One way to mitigate jamming attacks is using cognitive radio, a technology that enables radios to analyze frequencies and environmental conditions of their connectivity, allowing them to dynamically adapt transmission and reception characteristics, including frequency and signal strength. Using their capabilities, cognitive radios can detect and analyze the presence of signal interference caused by jammers. With this information, cognitive radios can autonomously and intelligently transfer routers or users to channels with minimal signal interference. By dynamically adjusting their operating frequencies and avoiding jammed channels, cognitive radios improve the resilience and robustness of the communication network against jamming attacks. This capability greatly contributes to ensuring uninterrupted and reliable communication in the face of deliberate signal interference~\cite{SALAMEH2020102035}.

Regulating the power of the transmitted signal is another effective strategy to prevent jamming attacks, as it poses challenges to attackers who attempt to disrupt communication. Using a high power signal, the attacker has difficulties executing a successful jamming since their interfering signal must surpass the strength of the original signal. By regulating the power of the transmitted signal, the system can establish a robust communication link that requires a higher power threshold to achieve successful jamming. Thus, an attacker needs to generate a stronger signal to overpower the regulated transmission, making the jamming attempt more challenging and resource-intensive. The power regulation technique serves as a countermeasure to deter and impede jamming attacks by requiring attackers to invest considerable resources to disrupt the communication. This approach strengthens the system's resilience and enhances its ability to sustain reliable communication even in the presence of potential jammers~\cite{9733393}.

\subsection{Sleep Deprivation}

The sleep deprivation attack~\cite{9902998} is a specific type of Denial-of-Service (DNS) attack that aims to disrupt device normal operation by causing prolonged wakefulness periods, leading to increased battery consumption and reduced device lifetime. This attack renders the device inoperable over time. In addition to targeting individual devices, the attacker can also exploit this attack to exhaust critical components within the IoT network, thereby affecting the overall stability and availability of the network. 

The attack focuses primarily on compromising device sensors, making it appear as if the devices are entering sleep mode when, in reality, their operational lifespan is being depleted. As a result, affected devices begin to experience malfunctions and contribute to network instability. The cumulative impact of these compromised devices can lead to significant disruptions in the availability of services. It disrupts the normal sleep-wake cycle of IoT devices, causing them to remain constantly active and consume excessive power. This attack strategy drains the device's battery and shortens its overall lifespan, rendering it eventually inoperable. In addition, the attack extends its impact beyond individual devices, affecting the stability and availability of the entire IoT network~\cite{jenifer2023detecting}.

\vspace{1em}
\textbf{Mitigations}

Several measures can be implemented to mitigate sleep deprivation attacks. Firstly, the deployment of strong authentication mechanisms~\cite{1495974} is crucial to ensure that only authorized devices can access the network. By verifying the identity of the devices before granting access, the risk of unauthorized devices engaging in sleep deprivation attacks can be significantly reduced. In addition, anomaly detection techniques~\cite{10024090} play a vital role in identifying abnormal behavior exhibited by devices. By continuously monitoring device activities and comparing them to established patterns, any deviations that may indicate a sleep deprivation attack can be quickly detected, allowing for timely intervention and appropriate response measures. Regular firmware updates, incorporating the latest security patches and enhancements, are essential to address known vulnerabilities and improve the resilience of IoT devices against sleep deprivation attacks. By ensuring that devices are running updated and secure firmware, the likelihood of successful attacks can be minimized. 

Continuous monitoring and device behavior analysis are critical to detecting and responding to sleep-deprived attacks~\cite{jenifer2023detecting}. Using real-time monitoring systems, abnormal patterns or prolonged activity that indicate a sleep deprivation attack can be quickly identified. In summary, mitigating sleep deprivation attacks requires a comprehensive approach that includes robust security protocols, effective monitoring systems, authentication mechanisms, anomaly detection, secure firmware updates, and proactive analysis of device behavior. Implementing these measures collectively strengthens the security posture of IoT networks and reduces their vulnerability to sleep-deprivation attacks, ensuring the continued stability and uninterrupted functioning of services.

\subsection{Physical capture of IoT devices}

The IoT system comprises a vast network of interconnected devices, each serving a specific function that collectively contributes to the overall operation of the system. However, the interconnected nature of these devices also makes the system susceptible to various hacker attacks. In some cases, attackers may directly target individual devices, requiring physical proximity to capture and compromise them in an attempt to extract valuable information. By gaining physical possession~\cite{7562568} of a device, attackers can attempt to infiltrate the network by impersonating a legitimate user. This unauthorized access allows them to exploit vulnerabilities within the network infrastructure and potentially gain access to sensitive content circulating within the system. 

With the device compromised and access to the network, the attacker can further exploit vulnerabilities to execute a range of additional attacks. These may include flooding the network with excessive traffic, disrupting the connectivity of legitimate devices, or launching DNS attacks to manipulate the system's domain name resolution. Through these attacks, the hacker can not only gain access to more devices within the network but also acquire sensitive information that can be exploited for malicious purposes. This escalating chain of attacks poses a severe threat to the security and integrity of the IoT system, potentially compromising user privacy, disrupting operations, and facilitating unauthorized access to critical resources~\cite{jsss.2022.07,GE201712}.

\vspace{1em}
\textbf{Mitigations}

To mitigate such attacks, proper physical security measures should be employed to protect the physical integrity of IoT devices, reducing the likelihood of unauthorized access or tampering. In addition, robust security measures must be implemented, including, but not limited to strong authentication mechanisms, encryption protocols, intrusion detection systems, and regular security updates. By adopting comprehensive security strategies, IoT systems can improve their resilience to attacks, ensuring the confidentiality, integrity, and availability of data and services~\cite{STELLIOS2021102316}.

Ensuring the physical security of these devices is crucial to prevent the capture of devices within the IoT environment. By implementing robust physical security measures, the attacker's ability to gain unauthorized access to the devices is significantly hindered. Various security means can be used to improve physical security and deter theft or unauthorized handling of IoT devices. 

\subsection{Phishing}

Phishing attacks~\cite{9290047} are deceptive schemes that frequently exploit social engineering tactics to deceive individuals and can manifest through various means, including deceptive emails that appear genuine and trustworthy. Attackers exploit vulnerabilities in the SMTP protocol, enabling them to send emails that may even appear to be from the recipient's address. These attacks can take various forms, such as fraudulent websites, deceptive emails, or other communication channels~\cite{Nmachi2021Phishing}. However, their underlying objective remains consistent: to illicitly acquire confidential information, including sensitive credentials, i.e., to deceive individuals and extract sensitive information~\cite{6497928}. By successfully executing a phishing attack, hackers can gain unauthorized access to networks or IoT devices, posing severe risks to the network and connected IoT devices. 

Phishing attacks are prevalent across all network types, including IoT networks such as 5G-IoT, as they exploit human vulnerabilities rather than specific technical weaknesses. In the context of 5G-IoT, if a phishing attack proves successful, the attacker can hide himself within the network, masquerading as a legitimate user~\cite{Alanazi20235GSecurity}. This attack allows them not only to steal confidential information but also to gain authorized access to the network and any connected IoT devices. The consequences of such an attack extend beyond the theft of personal information, as the attacker can leverage their authorized access to execute further attacks, compromising the integrity, confidentiality, and availability of the network and IoT devices.

\vspace{1em}
\textbf{Mitigations}

Preventive measures against phishing scams primarily involve scrutinizing the authenticity of the message and its source, as well as ensuring secure web browsing practices. It is crucial to exercise caution and verify the legitimacy of the messages received. Individuals should carefully analyze the content of emails, paying attention to any inconsistencies or suspicious elements. Verification of the sender's identity and the email's source can be done by contacting the supposed sender through alternative means, such as a phone call or official website, to confirm the authenticity of the communication~\cite{Alanazi20235GSecurity}.

Verification of website certificates is another valuable practice in preventing phishing attacks. Validating the authenticity of website certificates ensures that the communication with the website is encrypted, protecting sensitive information transmitted between the user and the website. It is essential to avoid providing confidential or personal data on websites lacking the HTTPS protocol, as HTTP websites lack the necessary security measures to safeguard such information adequately~\cite{6497928}.

\subsection{Spoofing}

A spoofing attack is a deceptive attack method that resembles phishing, as the attacker disguises himself as a trusted and authorized device or user to deceive and steal information~\cite{10.1016/j.comnet.2020.107658}. This type of attack can be executed through various techniques. In network-directed spoofing attacks, the adversary often employs IP spoofing, where they send messages with a falsified or masked IP address to make it appear as if they are from a trusted source~\cite{10.1109/IMSNA.2013.6743326}. In an IoT context, cybercriminals attempt to leverage spoofing attacks to spread malware throughout the network, potentially infecting multiple devices.

Several traditional forms of spoofing attacks exist, including: 
\begin{itemize}
    \item IP address spoofing - attackers forge the source IP address in network packets, making it appear as if they are sent from a different IP address than the actual source; 
    \item ARP spoofing - attackers associate their own MAC address with an authorized IP address already present on the network, intercepting and manipulating network traffic; 
    \item DNS spoofing - attackers exploit DNS vulnerabilities, employing techniques like cache poisoning to redirect network traffic intended for a specific domain name to an alternative IP address. 
\end{itemize}

Successful execution of a spoofing attack enables the attacker to impersonate a legitimate user within an IoT network. By doing so, they gain unauthorized access to restricted content and can exploit additional vulnerabilities to conduct more sophisticated attacks, such as DoS attacks, potentially infecting all or part of the system.

\vspace{1em}
\textbf{Mitigations}

Effective management of authentication keys plays a critical role in mitigating spoofing attacks. Key management techniques can be classified into symmetric, asymmetric, or group-based approaches~\cite{samiullah23_keys}. A widely used symmetric key management protocol is the Diffie-Hellman key exchange~\cite{gegenhuber24_diff}, which facilitates secure communication and the establishment of shared secret keys. Efficient and scalable key management becomes particularly important when dealing with numerous authentication instances within a system. 

Additionally, physical layer security provides an alternative solution to mitigate spoofing attacks~\cite{li21_phylayer}. This approach takes advantage of the inherent randomness and reciprocity of wireless channels to generate authentication keys. By exploiting the unique characteristics of wireless communication channels, such as signal strength, arrival time, and channel response, authentication keys can be securely established. The randomness of these properties of the wireless channels ensures the uniqueness and unpredictability of the authentication keys, enhancing the security of the communication system.

\subsection{Tampering}

Tampering attacks involve the unauthorized modification or replacement of device codes, confidential information, user access credentials, or other critical data. This attack shares similarities with eavesdropping, but goes beyond information theft by allowing the attacker to manipulate and modify data, directly impacting confidentiality and integrity~\cite{10.1109/COMST.2019.2910750}. Similarly to eavesdropping attacks, tampering attacks involve the interception of network communications, enabling the attacker to impersonate a legitimate user and engage in information exchange with other users or devices. This unauthorized access allows the attacker to obtain privileged data and manipulate the parameters exchanged within the IoT ecosystem. For instance, the attacker can modify information related to IoT devices or user access credentials, compromising the integrity and trustworthiness of the system~\cite{10.1109/ISCC.2015.7405513}.

Tampering attacks pose significant risks to the confidentiality, integrity, and availability of the IoT network. By tampering with data, attackers can exploit vulnerabilities in the system, potentially leading to unauthorized access, unauthorized modifications, or malicious manipulation of critical information~\cite{10.1109/ICGCIoT.2015.7380703}.

\vspace{1em}
\textbf{Mitigations} 
Hash codes provide an effective means to verify the integrity and detect tampering of files within an information system~\cite{wang23_hash}. A hash code is a unique digital fingerprint generated by a cryptographic hash function that transforms the contents of a file into a fixed-size alphanumeric string. Any alteration or modification made to the file results in a different hash code, enabling the identification of tampering attempts.

\subsection{Replay Attack}

Replay attacks represent a threat in which an attacker intercepts the content of a message and forwards it to a different destination, typically with the objective of gaining unauthorized access to a network or system~\cite{Elsaeidy20_replay,aqeel22_replay}. In essence, a replay attack involves the retransmission or relay of an authentication signal from a legitimate user to deceive the network into granting access. Replay attacks are classified as simple and easy to execute once the attacker does not necessarily need to decrypt the stolen message content. Instead, they focus on capturing and forwarding the message, taking advantage of the inherent trust in legitimate authentication signals or data exchanges.

\vspace{1em}
\textbf{Mitigations}

Single-use encryption methods are effective in mitigating replay attacks by ensuring that the forwarded message or code cannot be reused by the attacker~\cite{Abdelhafez23_repla,DERANGO20_replay}. By employing encryption techniques that generate unique encryption keys for each transmission, the security of the message is preserved even if it is intercepted and forwarded. When a single-use encryption method is implemented, the encryption key used to protect the message is unique and specific to that particular transmission. Once the message is received and decrypted by the intended recipient, the encryption key becomes obsolete and cannot be used to decrypt subsequent forwarded copies. This prevents the attacker from gaining access to the original message content or sensitive information. 

Furthermore, implementing a ``temporary'' message approach can improve security against replay attacks~\cite{DERANGO20_replay}. By incorporating an expiration mechanism within the message, the message's validity is limited to a specific timeframe. If the attacker attempts to access or forward the message after it has expired, the content becomes inaccessible or irrelevant, rendering the retransmission attempt futile.

\subsection{De-authentication}

The de-authentication attack can disrupt communication stability between devices within a network. This attack occurs when the attacker deliberately sends packets with the intention of disconnecting the target devices from the wireless network. By successfully disconnecting a device, the attacker gains the opportunity to send additional packets to disconnect more devices, which can result in system unavailability and instability~\cite{10.1109/ACCESS.2025.3623690}.

The success of a de-authentication attack relies on the ability to send packets that disrupt the normal operation of targeted devices, causing them to lose their wireless network connection. Once disconnected, the attacker can continue to target and disconnect more devices, potentially affecting a significant portion of the network. The impact of such an attack on an IoT ecosystem can be severe, as the system's functionality is heavily dependent on the interconnection and continuous communication between devices~\cite{10.1109/CONIT59222.2023.10205679}.

\vspace{1em}
\textbf{Mitigations}

One way of mitigating de-authentication attacks can be through the implementation of management systems that analyze the network traffic and control the rate of requests being made~\cite{10.1109/SMC.2015.55}. By effectively managing and filtering incoming requests, these systems can prevent the system from being overwhelmed by a high volume of malicious requests, rendering the de-authentication attack ineffective. 

Management systems play a crucial role in monitoring and analyzing network traffic patterns in real-time~\cite{10.1023/A:1009481719707}. They employ various techniques such as traffic analysis, anomaly detection, and rate limiting to identify and discard invalid or suspicious requests. By setting thresholds and rate limits, the management system can control the number of requests allowed within a specific timeframe, ensuring that the system remains operational and stable.

\subsection{Resource Exhaustion}

Hardware resource exhaustion attacks aim to deplete the capacity and resources of targeted devices~\cite{10.62019/abbdm.v4i4.261}. This attack employs various strategies to exhaust critical resources, impairing the device's functionality, and disrupting its common operation. For instance, in routers, resource exhaustion attacks specifically target the device's processing power, memory, and forwarding capacity in terms of packets per second. In the case of firewalls and intrusion detection systems (IDS), the attack focuses on preventing the establishment of new connections~\cite{10.1109/CNS53000.2021.9705038}.

Within an IoT network, hardware resource exhaustion attacks resemble a DoS attack, where the attacker inundates the user or device with numerous requests, overwhelming the system's available resources. The primary objective is to exhaust key system resources, such as battery power and memory. The impact of such an attack can extend beyond the targeted device and adversely affect the entire network, resulting in network unavailability and potential damage to the devices themselves~\cite{10.1007/978-3-319-66379-1_23}.

\vspace{1em}
\textbf{Mitigations}

Monitoring software serves as a valuable tool for detecting resource exhaustion attacks, as it enables the observation of increased processing power demands that indicate that such attacks are taking place. Continuous monitoring of system performance metrics, such as CPU utilization, memory usage, and network traffic, enables identifying patterns or deviations that indicate resource exhaustion~\cite{10.3390/info12040154}. Indicators of Attack (IoA) can also provide valuable insight into resource exhaustion and help prevent such attacks within an IoT network. IoAs are predefined patterns or behaviors associated with known attack techniques. By analyzing network traffic, system logs, and other relevant data, IoAs can identify suspicious activities that indicate the presence of a resource exhaustion attack~\cite{10.3390/fi17050226}.

\subsection{Blueborne}

Blueborne is a significant attack vector that leverages Bluetooth connections to compromise devices, gain unauthorized access to corporate data and networks, infiltrate secure networks, and propagate malware to other devices~\cite{almiani19_blue,mann20_blue}. This attack vector poses a significant threat, as it exploits vulnerabilities in about $25\%$ of IoT devices~\cite{sinhaIoTlytcs_25}, as virtually all devices equipped with Bluetooth are susceptible to Blueborne attacks. Blueborne acts as a gateway for other malicious activities, enabling attackers to exploit the initial vulnerability and subsequently launch spying attacks, engage in data theft, or execute other malicious activities. This attack vector poses a severe risk to the security and privacy of IoT devices, as well as the networks to which they are connected.

The emergence of 5G technology, which is widely connected to various IoT devices, introduces additional gateways for Blueborne to infiltrate the 5G network, potentially impacting a larger number of devices.

\vspace{1em}
\textbf{Mitigations} 

Network security solutions and firewalls focus primarily on preventing threats that spread through IP-based networks. However, these traditional protections may not be as effective in mitigating the risks posed by Blueborne attacks, as Blueborne exploits vulnerabilities in Bluetooth communication, which occurs over the airwaves rather than through IP-based protocols. Given the nature of Blueborne attacks, combating this threat requires specialized solutions that specifically target the vulnerabilities associated with Bluetooth connections. Implementing a user-space script on the Bluetooth
master to flip the status of sub-channels for hopping can make it harder for a sniffer to learn the adaptive
hopping technique~\cite{10.1145/2906388.2906403}, and creating a fresh permanent key for each session once a new Bluetooth device is added to the piconet, are possible mitigation strategies~\cite{10.1145/3395351.3399343}.

A key objective is to develop effective countermeasures that inhibit Blueborne infiltration into secure networks, in addition to keeping devices updated, thus preventing device compromise and potential malware spread~\cite{jsan18blue}. Other users' actions to mitigate Blueborne attacks include turning off Bluetooth when it is not in use, using a long PIN for the authentication phase, removing, unpairing, and deleting access to devices that are lost or stolen, and being aware of all of the Bluetooth devices that are connected in their piconet~\cite{10.1007/978-3-030-93956-4_7}.

\subsection{Man-In-The-Middle}

The Man-In-The-Middle (MITM) attack is a well-known and prevalent attack in which an attacker stands between the communication of two legitimate nodes and intercepts all the transmitted data. By positioning himself as a middle point that intercepts data, the attacker deceives the nodes, making them believe that communication is taking place with a trusted and authorized entity. This enables the attacker to gain unauthorized access to transmitted information, potentially even modifying or preventing it from reaching its intended destination~\cite{10.5267/j.ijdns.2019.1.001}. One of the key challenges in detecting an MITM attack is that the targeted nodes are unable to identify the presence of the fraudulent node. As a result, they remain unaware that their messages are being intercepted and manipulated by the attacker. This inherent difficulty in recognizing the attacker's presence complicates identifying and countering the attack within the network~\cite{alaba2017}.

Once successfully established a man-in-the-middle position, the attacker can exploit this privileged position to launch additional, more severe attacks against the targeted devices. For instance, the attacker may hijack valid sessions to gain unauthorized access to sensitive information, carry out DoS attacks to disrupt the normal operation of the targeted devices, or redirect the victim to a fake address, potentially leading to further exploitation~\cite{10.1109/COMST.2016.2548426}.

\vspace{1em}
\textbf{Mitigations}

Mutual authentication techniques are widely used to establish secure communication channels and prevent data interception by malicious entities that pose as legitimate stations~\cite{SHAMSHAD_22+mitmprev}. These techniques ensure that both communicating entities, such as clients and servers, authenticate each other's identities before exchanging sensitive information. By requiring mutual authentication, the risk of MITM attacks is significantly reduced. Passive detection systems represent another effective solution to detect and mitigate MITM attacks. These systems continuously monitor network traffic, analyzing packets transmitted across the network~\cite{10.1109/ACCESS.2024.3362803}. When abnormal traffic patterns or suspicious behavior are detected, such as the presence of unauthorized entities or unexpected communication patterns, the passive detection system issues an attack alert, allowing administrators to respond promptly and mitigate the attack. 

Another possibility to mitigate MITM attacks is the use of centralized servers and the creation of comprehensive address lists to enhance the security of communication channels~\cite{kumar_tapaswi12_mitm_prev}. When devices establish communication, they can consult this list to verify the authenticity of the addresses of the devices they are interacting with. This centralized server model provides an additional layer of security against MITM attacks by enabling devices to independently verify the authenticity of their communication partners. Thus, mutual authentication techniques, passive detection systems, and decentralized server models can significantly enhance the security of communication channels and mitigate the risks posed by MITM attacks.

\subsection{Identity Theft}

The continued usage of default login credentials presents a serious security vulnerability, as attackers are well aware of these preset configurations and can easily exploit them to gain unauthorized access to devices~\cite{Knieriem2018DefaultPasswords,Alghamdi2025CCTV}. This issue enables attackers to bypass authentication mechanisms and compromise device security, as well as access sensitive data stored within the system. In addition to exploiting default credentials, attackers may also employ various techniques, such as deploying malware, to obtain user login information illegally. By infecting devices with malicious software, they can monitor user activity, capture keystrokes, and perform other covert actions to steal credentials without the user’s awareness~\cite{ahmed2014survey,Sharma2023KeystrokeLogger}. With this information, attackers can gain control of the device or exfiltrate sensitive data, resulting in privacy violations and other malicious activities. 

This type of attack does not necessarily rely on a compromised database or advanced hacking techniques~\cite{Vidalis2014IdentityTheft,e25050717}. Attackers can leverage social engineering strategies or phishing campaigns to trick users into voluntarily disclosing their login credentials~\cite{jcs.2023.041095}. Social engineering techniques exploit human behavior, trust, or emotions to persuade people to reveal sensitive information, such as usernames and passwords. Phishing attacks, in particular, often involve sending deceptive emails or messages that appear legitimate, luring users to provide their credentials or accessing malicious websites designed to harvest this information~\cite{Alkhalil2021Phishing}.

\vspace{1em}
\textbf{Mitigations}

The widespread use of default usernames and passwords in IoT devices poses a significant security risk, underscoring the need for immediate replacement of these credentials to enhance device protection and reduce the likelihood of unauthorized access. Substituting default credentials with strong and unique passwords is essential to prevent hacking attempts. By creating complex passwords that incorporate a mix of letters, numbers, and special characters, users can significantly enhance the security of their devices, making them more resistant to brute-force attacks based on automated guessing techniques~\cite{Saputra2025PasswordStrength}. 

To further reduce the risk of credential theft, users should remain vigilant and adopt proactive measures to protect their information. This involves avoiding the sharing of sensitive data, such as user credentials, with unauthorized individuals or untrusted websites~\cite{Kuraku2024CyberAwareness}. Building awareness of common social engineering tactics and phishing schemes is also crucial, as attackers frequently use these deceptive methods to manipulate users into revealing confidential information. Users should critically assess and verify the authenticity of any request for personal data, maintain a cautious attitude, and ensure the legitimacy of websites and communication channels before revealing sensitive information~\cite{Kudalkar2024PhishingAwareness}.

\subsection{Backdoor}

Backdoors are techniques designed to bypass authentication and other security mechanisms, allowing attackers to gain unauthorized access to a computer system or the data stored within it~\cite{Wysopal2010Backdoors}. These malicious actions are commonly associated with malware that seeks to gain control over the targeted system. Backdoor attacks often focus on systems with remote access capabilities, as these provide opportunities for attackers to intercept web traffic and capture sensitive data during transmission~\cite{chathoth2025pcapbackdoorbackdoorpoisoninggenerator}. The attack typically involves the installation of malicious code or the injection of harmful data, allowing the attacker to infiltrate the network without the user’s knowledge. To avoid detection, attackers employ various tactics to conceal their activities, making it difficult for users to notice anomalies or realize that their system has been compromised~\cite{10529225}. Once the backdoor is successfully established, all information exchanged within the affected system can be monitored and collected by the attacker, resulting in unauthorized access to sensitive information~\cite{9711191,khan2024backdoor}.

\vspace{1em}
\textbf{Mitigations}

Ensuring a proper firewall configuration is essential to prevent unauthorized access and reduce the risk of backdoor attacks~\cite{lindqvist2017future}. Firewalls play a critical role in network security by controlling incoming and outgoing traffic and enforcing access control policies. However, if not configured correctly, firewalls can unintentionally leave ports open or allow unauthorized connections, creating opportunities for attackers to establish backdoors. In addition to firewalls, the implementation of an Intrusion Prevention System (IPS) can further strengthen defenses against backdoor threats. An IPS continuously monitors network traffic in real time, actively detecting and blocking patterns or signatures associated with known malicious activities~\cite{asi8020052}.

\subsection{Sinkhole}

Attackers launch a sinkhole attack by deploying a deceptive node within the network, positioning it between the base station and neighboring nodes~\cite{Mehta2022Sinkhole}. They manipulate and control network traffic by broadcasting false routing information to attract packets to their malicious node. This strategy allows attackers to gain unauthorized access to network traffic that passes through the compromised nodes.

The sinkhole attack offers a significant advantage by easily combining with other attacks to amplify its impact. For example, attackers attract network packets to their malicious node by falsely advertising a more efficient routing path. Once the packets reach their node, the attackers selectively forward or drop specific packets based on their objectives~\cite{Rehman2019Sinkhole}. This selective forwarding enables them to manipulate the network data flow, potentially causing denial of service by discarding or delaying critical packets.

\vspace{1em}
\textbf{Mitigations}

Sinkhole attacks can be mitigated through strategies that combine preventive practices and specialized tools. Traffic monitoring mechanisms can be useful to prevent this attack, as it is possible to see unnecessary or suspicious changes in records~\cite{10562222}. Implementing strong authentication between devices on the network, such as digital certificates or cryptographic key-based authentication, is essential to ensure that only trusted nodes participate in communication~\cite{10275916,Attkan2022Cyber}. In addition, adopting secure routing protocols makes it more difficult for malicious nodes to insert themselves into the network topology~\cite{SOPHIEMARIAVINCENT2024123765,electronics12030482}. 

Another effective approach relies on continuous network analysis and monitoring. Anomaly detection tools, whether based on machine learning or rule-based systems, identify unusual traffic patterns that may signal a sinkhole attack~\cite{PRATHAPCHANDRAN2021108413,Giri2023Sinkhole}. Advanced firewalls and intrusion prevention systems (IPS) block suspicious traffic before it compromises network integrity. Additionally, network segmentation and traffic isolation policies, such as VLANs or software-defined networking, limit potential damage by preventing the spread of compromised data~\cite{10.1145/3737875}.

\subsection{Sybil}

In a Sybil attack, the adversary generates multiple false identities, known as Sybil nodes, within the network~\cite{PU2022102541}. These nodes impersonate legitimate and trustworthy devices to deceive other network participants. The attacker’s main goal is to gain unauthorized access to sensitive information stored by other nodes. During the attack, the adversary compromises the sensors or nodes and assigns false and fraudulent identities~\cite{10073944}. The attacker then uses these fake identities to perform malicious actions, such as eavesdropping on communications, injecting false data, or disrupting the normal operations of the network.

Sybil attacks often succeed because sensors typically operate with limited resources, including low memory capacity~\cite{10.1145/984622.984660}. This constraint prevents them from implementing a robust authentication and encryption mechanisms needed to verify the authenticity of other nodes on the network. As a result, attackers can easily exploit this vulnerability and launch Sybil attacks by bypassing proper identity verification.

\vspace{1em}
\textbf{Mitigations}

Validation techniques support mitigating Sybil attacks within a network. These techniques rely on a central entity that validates the identities exchanged between local and remote nodes~\cite{https://doi.org/10.1155/2016/9783072}. Acting as a trusted authority, the central entity ensures the authenticity and legitimacy of all participating identities~\cite{Levine2006a}. There are two main validation approaches: direct and indirect validation. In direct validation, the local node requests the central entity to verify the remote identity. In indirect validation, the local node uses previously verified identities to assess the legitimacy of the remote node. By establishing a network of trusted identities, the local node ensures that its communication partners are legitimate and not controlled by Sybil attackers.

\subsection{Masquerade Attacks}

In a masquerade attack, the attacker gains unauthorized access to a system or device by using forged or spoofed credentials~\cite{wardega2019resilience}. They exploit weak or missing authentication mechanisms to go undetected through the network and impersonate legitimate users or nodes~\cite{ustun2019novel}. To launch a successful attack, the attacker typically steals the victim’s credentials or installs malicious software, such as keyloggers or rootkits. These tools enable the attacker to collect sensitive information and gain control of the victim’s account or system. In some cases, careless user behavior (e.g., leaving a device connected to the network without proper security measures) provides the attacker with an opportunity to breach the system.

\vspace{1em}
\textbf{Mitigations}

To reduce the risk of masquerade attacks, organizations must implement strong authentication protocols that block unauthorized access through false credentials~\cite{10.1145/3703444,s24247967}. They should enforce robust password policies and avoid using default or easily guessable passwords. Adding multi-factor authentication, such as biometric verification or one-time passwords—strengthens security by introducing an extra layer of protection. In wireless sensor networks, effective deployment techniques help detect disguised attackers and significantly improve network security~\cite{bhuse2007detection,Sujihelen2022}.

\subsection{Skimming}

Attackers use skimming to clone the credentials of legitimate users~\cite{10.5555/1971859.1971863,sujatha2025atm}. In this method, they place a device—called a skimmer—near a card reader or access control system to secretly capture and copy physical credentials, such as ID cards or chips. These devices collect the necessary information without the user’s knowledge. Because skimming requires physical proximity, attackers typically carry out these attacks in public locations, such as ATMs, payment terminals, or secure entry points~\cite{mahammad2021internet}.

\vspace{1em}
\textbf{Mitigations}

To reduce the risk of skimming attacks, users must remain vigilant and take preventive measures when using devices that require personal credentials~\cite{al2015detecting,raman2023iot}. An effective practice is to carefully inspect the card reader or chip slot before inserting a card or credential. This means visually inspecting the device for signs of tampering, unusual materials, or suspicious attachments that could indicate a hidden skimming device.

\subsection{Insider Attack}

The insider threat occurs when an internal actor, intentionally or unintentionally, alters the configuration of a device, infrastructure, or other components of the system, creating potential vulnerabilities and exposing sensitive data~\cite{AMIRIZARANDI2023100965}. This type of threat can cause more damage than external attacks because the insider already has legitimate access and does not need to bypass authentication mechanisms. Although monitoring systems can effectively detect external threats, identifying insider attacks is more challenging, as they often originate from trusted users and occur without warning~\cite{ambili2019trust}.

\vspace{1em}
\textbf{Mitigations}

Mitigating insider attacks presents unique challenges that differ from those of detecting external threats. An effective approach involves monitoring host integrity and user activity to identify and respond to insider threats~\cite{Salem2008_insider,Alsowail_alshehari_21_insider,INAYAT24_insider}. By analyzing network behavior, the system can detect unusual changes in log files, system resources, and files. This monitoring allows the system to block further access from users who exhibit suspicious behavior. Preventing insider threats also depends on building user awareness and resilience, as many insider attacks result from unintentional actions. Educating users about risks such as phishing and social engineering and fostering a culture of security through ongoing training help reduce the likelihood of these threats.

\subsection{Node Replication}

In a node replication attack, the attacker captures a legitimate node on the network and creates multiple identical copies of it~\cite{mbarek2023effective}. They reintroduce these cloned nodes into the network, often without being detected. These replicated nodes can disrupt normal operations by overwhelming the network infrastructure, leading to service unavailability, performance degradation, congestion, and slower response times. Furthermore, node replication poses a threat to data integrity and confidentiality, as the attacker can use cloned nodes to intercept and manipulate sensitive information exchanged across the network~\cite{lee2018mdsclone}. Unauthorized access can result in data breaches, altered communications, and the exploitation of protected network resources.

\vspace{1em}
\textbf{Mitigations}

To prevent node replication in a network, forming subgroups with designated leader nodes and deploying a monitoring system are effective strategies to avoid node replication~\cite{lee2018monitoring}. For example, subgroups can organize the network into smaller clusters, with each leader node responsible for gathering and transmitting information within its group~\cite{abujassar2024enhancing}. This structure improves network integrity and simplifies oversight. In this approach, sub-trees within the network actively communicate and intersect to detect cloned nodes. By comparing node identifiers between subgroups, the system can identify duplicates, which are clear signs of replication. This detection mechanism may prevent the network from accepting and integrating cloned nodes.

\subsection{Wormhole}
In a wormhole attack, malicious actors take control of two or more network nodes to create a covert channel for unauthorized data transfer~\cite{abdullah51detecting}. This type of attack poses serious security risks by allowing attackers to intercept confidential information, launch denial-of-service attacks, and compromise network integrity. The main goal of attackers is to break the confidentiality of the data by tunneling packets through the hidden channel~\cite{bhosale2022wormhole}. This tunnel allows them to capture sensitive data exchanged between legitimate nodes, which could result in the loss of critical information. 

This attack also disrupts normal network operations by blocking data packets from reaching their intended destinations~\cite{chen2022novel}. The disruption causes instability, leading to service interruptions and a decrease in overall network performance~\cite{DESHMUKHBHOSALE2019840}. Detecting wormhole attacks is particularly challenging because attackers often use encapsulation techniques to hide packet contents. By concealing this information, they evade standard security measures, making it harder for traditional defenses to detect and stop the attack.

\vspace{1em}
\textbf{Mitigations}

A common strategy used to mitigate the risk of denial-of-service attacks, particularly in the context of wormhole attacks, is to implement packet transmission limits and distance restrictions for nodes within a network~\cite{SHAHID20211967}. By imposing limits on the number of packets that a node can transmit, network administrators can prevent the nodes from overwhelming the network with excessive traffic.

\subsection{Hijacking}
Hijacking attacks use various techniques to gain unauthorized control over communications, systems, or applications~\cite{mutleg2024comprehensive}. Each type of hijacking targets a specific part of the system. For example, browser hijacking takes over a user’s web browser, while DNS hijacking redirects traffic by controlling a domain name server~\cite{wedman2013analytical}. Aircraft hijacking involves seizing control of vehicles or aircrafts~\cite{10592010}, and session hijacking captures a legitimate user’s session during the login process.

In IoT networks, hijacking attacks pose a serious risk by allowing attackers to take control of legitimate connected devices. Once hijackers gain access, they can use these compromised devices for malicious purposes. For example, they might convert devices into bots to steal sensitive information from the system~\cite{al2020siem}. They can also flood the network with traffic, launching a distributed denial-of-service attack~\cite{s24113571}. These actions disrupt the regular operation of the IoT network, leading to instability and reduced service availability.

\vspace{1em}
\textbf{Mitigations}

To reduce the risk of hijacking attacks on IoT devices, it is crucial to address the common use of weak or default passwords~\cite{Alghamdi2025CCTV}. Enforcing strong passwords that are difficult to guess helps protect devices from brute-force attacks. Adding multi-factor authentication strengthens security by requiring users to verify their identity in multiple ways. Using virtual private networks further enhances protection by creating secure encrypted tunnels between devices and the network~\cite{goyzueta2021vpnot}. These tunnels prevent unauthorized access and block attempts at eavesdropping.

The deployment of an access control management system is another effective way to secure IoT devices~\cite{singh2023access}. This type of system enables administrators to define and enforce access rules, ensuring that only authorized users or devices can access the network. By strictly managing user permissions, administrators can significantly reduce the risk of unauthorized access and hijacking. It is also essential to protect the session ID, which acts as an authentication token during user sessions. Sharing this ID only with trusted sources helps prevent session hijacking~\cite{humaira2020secure}. Additionally, using secure protocols (e.g., HTTPS) and keeping device firmware and software up-to-date further enhances IoT security.

\subsection{Hello Flooding}

The Hello Flood is a disruptive network attack that severely degrades network performance~\cite{electronics13112226,koul2023detection}. In this attack, a malicious node is introduced into the network and floods neighboring nodes with an excessive number of HELLO messages. Unlike many other attacks, the Hello Flood does not require the attacker to forge authentication credentials. Instead, by broadcasting recognition messages that appear legitimate, the attacker deceives nearby nodes into trusting it as a genuine part of the network.

This attack, classified as a routing or DoS attack, exploits the trust that nodes establish with one another~\cite{maurya2022impact}. Legitimate nodes mistakenly identify the malicious node as a trustworthy participant and begin routing packets and other traffic toward it. This misdirection leads to network congestion, which degrades both performance and availability. Beyond congestion, the Hello Flood attack also drains network resources. The nodes located near the attacker expend more energy and processing power than usual, resulting in higher power consumption and accelerated resource depletion.

\vspace{1em}
\textbf{Mitigations}

Monitoring software is crucial for detecting and mitigating Hello Flood attacks by identifying abnormal network behavior in real time~\cite{rani2025semi}. By continuously analyzing network traffic and activity, the software can recognize suspicious patterns, such as an unusually high number of HELLO messages sent from a single node, that signal a potential attack. In addition to monitoring tools, protocols with built-in bidirectional checks provide an extra layer of defense~\cite{saghar2015raeed}. These protocols require nodes to establish two-way communication, allowing them to verify the legitimacy of their neighbors before trusting them.

\subsection{Desynchronization Attack}
\label{subsec:desync}
A desynchronization attack is a form of DoS attack that disrupts the authentication process within a network~\cite{aghili2018impersonation,mokhtari2025practical}. In this attack, the attacker intentionally breaks the synchronization between the clients and the access point, preventing legitimate users from successfully authenticating and accessing the network. The attacker manipulates the authentication requests exchanged between the client and the access point, interfering with their communication. By altering these requests, the attacker disrupts the authentication process, rendering legitimate credentials ineffective and blocking client access. 

It is essential to distinguish a desynchronization attack from other similar threats, such as man-in-the-middle or disconnection attacks~\cite{salem2021man,bernardinetti2020disconnection}. Although it shares specific characteristics with these attacks, its main goal is to prevent users from successfully authenticating with the network. Unlike a man-in-the-middle attack, the attacker does not intercept or relay communications; instead, they focus on disrupting the authentication process by manipulating authentication requests. Similarly, unlike a disconnection attack, which only temporarily disconnects devices, the desynchronization attack prevents the device from completing the login process altogether.

\vspace{1em}
\textbf{Mitigations} 

Two-factor authentication (2FA) and encryption are effective strategies for strengthening network security and preventing unauthorized access, especially in the context of desynchronization attacks~\cite{sym16101282}. These methods add layers of verification and data protection, making it more difficult for attackers to interfere with the authentication process~\cite{gope2018lightweight}. Additionally, using the most secure communication protocols is essential to reduce the risk of desynchronization attacks by ensuring the integrity and confidentiality of authentication exchanges.

\subsection{Grayhole/Blackhole}
Selective Forwarding (also called Grayhole) and Blackhole attacks disrupt network communications by manipulating how packets are forwarded~\cite{Alansari2023RPLAD3}. In a Selective Forwarding attack, the attacker intentionally drops certain packets while forwarding others to specific nodes. By doing this, the attacker disrupts the normal flow of communication and can cause a denial-of-service attack by preventing the delivery of critical packets. In a Blackhole attack, the attacker discards all received packets, completely blocking communication through the compromised node~\cite{doi:10.3233/AIS-210591}. This attack is usually easier for monitoring software or network administrators to detect because the sudden absence of packet flow is an abnormal behavior in the network.

\vspace{1em}
\textbf{Mitigations}

Intrusion detection systems can identify and mitigate attacks in network environments~\cite{kurtkoti2022performance,li2024cooperative}. These systems can detect and manage various types of attack, including those that exploit network density, node mobility, or group data forwarding. Another approach uses watchdog mechanisms, where nodes actively monitor their neighbors to verify packet forwarding~\cite{Huang2008ExtendedWatchdog}. If a node consistently drops packets, the system flags it as suspicious. However, collisions or mobility can affect this method, so it is often complemented by path redundancy, which sends data over multiple routes to ensure delivery even if one path is compromised.

\begin{table*}[!tbh]
\centering
\caption{Summary of IoT Attacks and Mitigation Strategies}
\label{tab:att_mit_summary}
\renewcommand{\arraystretch}{1.2} 
\begin{tabular}{|l|l|p{8.5cm}|}
\hline
\textbf{Attack Name} & \textbf{Main IoT Layer} & \textbf{Primary Mitigation(s)} \\ \hline
Eavesdropping & Network/Communication & Secure transmission channels, artificial noise, encoding, and cryptography.  \\ \hline
Brute-Force & Application/Service & Strong unique passwords, avoiding default credentials, 2FA/MFA, and zero-trust models. \\ \hline
DDoS & Application/Service & Network monitoring, access control, traffic filtering, and system resilience measures.  \\ \hline
Jamming & Network/Communication & Cognitive radio for dynamic channel switching and transmitted signal power regulation. \\ \hline
Sleep Deprivation & Perception/Physical & Strong authentication, anomaly detection, secure firmware updates, and continuous monitoring. \\ \hline
Physical Capture & Perception/Physical & Physical security measures, robust authentication, IDS, and encryption.  \\ \hline
Phishing & Application/Service & Verification of message authenticity, secure web browsing, and HTTPS certificate validation. \\ \hline
Spoofing & Network/Communication & Symmetric/asymmetric key management (e.g., Diffie-Hellman) and physical layer security.  \\ \hline
Tampering & Network/Communication & Hash codes and cryptographic hash functions for data integrity verification.  \\ \hline
Replay Attack & Network/Communication & Single-use encryption methods and temporary messages with expiration mechanisms. \\ \hline
De-authentication & Network/Communication & Management systems for real-time traffic analysis, rate limiting, and anomaly detection. \\ \hline
Resource Exhaustion & Perception/Physical & Continuous monitoring software (CPU/memory) and Indicators of Attack (IoA) analysis. \\ \hline
Blueborne & Network/Communication & Sub-channel hopping scripts, fresh permanent keys per session, and long PINs. \\ \hline
Man-In-The-Middle & Network/Communication & Mutual authentication, passive detection systems, and centralized address lists. \\ \hline
Identity Theft & Application/Service & Replacement of default credentials and user awareness against social engineering. \\ \hline
Backdoor & Perception/Physical & Proper firewall configurations and Intrusion Prevention Systems (IPS). \\ \hline
Sinkhole & Network/Communication & Traffic monitoring, strong authentication, secure routing, and anomaly detection (ML). \\ \hline
Sybil & Perception/Physical & Direct and indirect identity validation techniques via central trusted entities. \\ \hline
Masquerade Attack & Application/Service & Robust password policies, multi-factor authentication (MFA), and biometric verification. \\ \hline
Skimming & Perception/Physical & Vigilance and visual inspection of card readers for tampering or suspicious attachments. \\ \hline
Insider Attack & Application/Service & Monitoring host integrity and user activity logs, and cybersecurity awareness training.  \\ \hline
Node Replication & Perception/Physical & Network sub-grouping with leader nodes and node identifier comparison/monitoring.  \\ \hline
Wormhole & Network/Communication & Implementing packet transmission limits and distance restrictions for network nodes. \\ \hline
Hijacking & Application/Service & Strong passwords, MFA, VPNs, secure session IDs, and access control management. \\ \hline
Hello Flooding & Network/Communication & Real-time traffic monitoring and protocols with built-in bidirectional checks.  \\ \hline
Desynchronization & Network/Communication & Two-factor authentication (2FA), encryption, and secure communication protocols. \\ \hline
Grayhole/Blackhole & Network/Communication & IDS, watchdog mechanisms, path redundancy, trust/reputation systems, and secure routing. \\ \hline
Linkability Attack & Network/Communication & Pseudonym rotation, Mix networks, onion routing, traffic shaping, and timing obfuscation. \\ \hline
\end{tabular}
\end{table*}

Another effective mitigation strategy involves implementing trust and reputation systems~\cite{diao2023trusted,borges2025self}. In these systems, nodes continuously evaluate the behavior of their neighbors and assign trust scores based on the level of cooperation they exhibit. The network can then avoid using nodes with low scores in routing decisions. In addition, authentication processes and secure routing protocols, such as SAODV or ARAN, help ensure that only trusted nodes can participate in the network, reducing the risk of malicious node insertion~\cite{10.1145/3586102.3586128,nugraha2017performance}.

Moreover, researchers have turned to machine learning approaches to detect anomalous behavior in network traffic that can signal grayhole or blackhole attacks~\cite{ioulianou2022ml}. By training models on patterns of normal and malicious behavior, these systems can identify attacks in real time with high accuracy. Combining these techniques, such as monitoring, trust management, secure routing, and intelligent detection, creates a robust defense against grayhole and blackhole attacks in dynamic and decentralized wireless environments.

\subsection{Linkability Attack}

Linkability attacks pose a privacy threat by allowing an adversary to connect two or more pieces of information or communications to the same source or user, even if the user remains anonymous~\cite{xu2024anti,song2020analyzing}. In wireless networks - especially in the Internet of Things (IoT) or mobile environments - attackers exploit patterns such as message timing, device identifiers (like MAC addresses), signal characteristics, or routing paths to associate different transmissions with a specific device or individual. Although the user's actual identity may not be directly exposed, this ability to correlate activities can result in serious privacy violations, including user profiling, tracking, or behavior analysis.

\vspace{1em}
\textbf{Mitigations}

To mitigate linkability attacks, networks can adopt several techniques, such as pseudonym rotation or identifier randomization, to prevent attackers from exploiting static identifiers~\cite{liu2019uncoordinated,wang2023privacy}. Mix networks and onion routing help obscure communication patterns by sending messages through multiple intermediate nodes~\cite{jambha2024securing}. In addition, traffic shaping and timing obfuscation techniques reduce the predictability of transmissions, making them harder to correlate~\cite{worae2025hiding,10.1145/3229565.3229567}.

\subsection{Summary of Attacks and Mitigation Strategies}

To provide a consolidated view of the threats discussed throughout this section, Table \ref{tab:att_mit_summary} summarizes the 28 identified IoT attacks alongside their corresponding architectural layers and primary mitigation strategies. The attacks are distributed across the Perception/Physical, Network/Communication, and Application/Service layers, reflecting the diverse technical entry points exploited by malicious actors. The table highlights that mitigation strategies vary significantly depending on the affected layer: while some threats require physical security enhancements or lightweight cryptography at the device level, others demand robust network monitoring, machine learning-based anomaly detection, or strict access control mechanisms at the application level. By aggregating these mitigation approaches and their key references, this summary serves as a quick reference guide for researchers and practitioners aiming to deploy targeted defense mechanisms tailored to the specific vulnerabilities of their IoT ecosystems.

%% file: StrideCVSS.tex
\section{STRIDE and CVSS Classification}
\label{sec:stridecvss}

\definecolor{Alto}{rgb}{0.85,0.85,0.85}
\definecolor{BlazeOrange}{rgb}{1,0.427,0.003}
\definecolor{SelectiveYellow}{rgb}{0.984,0.737,0.015}
\definecolor{VeniceBlue}{rgb}{0.043,0.325,0.58}

\begin{table}[t]
\centering
\caption{Attack classification using STRIDE and CVSS}
\label{tab:stridecvss}
\begin{tabular}{|l|c|c|c|c|c|c|} 
\hline
\textbf{Attack} & \textbf{S} & \textbf{T} & \textbf{R} & \textbf{I} & \textbf{D} & \textbf{E} \\
\hline
Eavesdropping & \cellcolor{Alto} & \cellcolor{Alto} & \cellcolor{Alto} & \cellcolor{BlazeOrange} & \cellcolor{Alto} & \cellcolor{Alto} \\
\hline
Brute-Force & \cellcolor{SelectiveYellow} & \cellcolor{Alto} & \cellcolor{Alto} & \cellcolor{Alto} & \cellcolor{Alto} & \cellcolor{Alto} \\
\hline
Distributed Denial of Service (DDoS) & \cellcolor{Alto} & \cellcolor{Alto} & \cellcolor{Alto} & \cellcolor{Alto} & \cellcolor{BlazeOrange} & \cellcolor{Alto} \\
\hline
Jamming & \cellcolor{Alto} & \cellcolor{Alto} & \cellcolor{Alto} & \cellcolor{BlazeOrange} & \cellcolor{BlazeOrange} & \cellcolor{Alto} \\
\hline
Sleep Deprivation & \cellcolor{Alto} & \cellcolor{Alto} & \cellcolor{Alto} & \cellcolor{Alto} & \cellcolor{BlazeOrange} & \cellcolor{Alto} \\
\hline
Physical capture of IoT devices & \cellcolor{Alto} & \cellcolor{Alto} & \cellcolor{Alto} & \cellcolor{SelectiveYellow} & \cellcolor{SelectiveYellow} & \cellcolor{Alto} \\
\hline
Phishing & \cellcolor{SelectiveYellow} & \cellcolor{Alto} & \cellcolor{SelectiveYellow} & \cellcolor{SelectiveYellow} & \cellcolor{Alto} & \cellcolor{Alto} \\
\hline
Spoofing & \cellcolor{BlazeOrange} & \cellcolor{VeniceBlue} & \cellcolor{Alto} & \cellcolor{BlazeOrange} & \cellcolor{VeniceBlue} & \cellcolor{Alto} \\
\hline
Tampering & \cellcolor{BlazeOrange} & \cellcolor{VeniceBlue} & \cellcolor{SelectiveYellow} & \cellcolor{BlazeOrange} & \cellcolor{Alto} & \cellcolor{Alto} \\
\hline
Replay & \cellcolor{SelectiveYellow} & \cellcolor{Alto} & \cellcolor{SelectiveYellow} & \cellcolor{SelectiveYellow} & \cellcolor{Alto} & \cellcolor{Alto} \\
\hline
De-authentication & \cellcolor{Alto} & \cellcolor{Alto} & \cellcolor{Alto} & \cellcolor{Alto} & \cellcolor{BlazeOrange} & \cellcolor{Alto} \\
\hline
Resource Exhaustion & \cellcolor{Alto} & \cellcolor{Alto} & \cellcolor{Alto} & \cellcolor{Alto} & \cellcolor{BlazeOrange} & \cellcolor{Alto} \\
\hline
Blueborne & \cellcolor{Alto} & \cellcolor{BlazeOrange} & \cellcolor{Alto} & \cellcolor{SelectiveYellow} & \cellcolor{BlazeOrange} & \cellcolor{Alto} \\
\hline
\textit{Man-In-The-Middle} & \cellcolor{SelectiveYellow} & \cellcolor{BlazeOrange} & \cellcolor{SelectiveYellow} & \cellcolor{BlazeOrange} & \cellcolor{Alto} & \cellcolor{SelectiveYellow} \\
\hline
Identity Theft & \cellcolor{BlazeOrange} & \cellcolor{BlazeOrange} & \cellcolor{SelectiveYellow} & \cellcolor{BlazeOrange} & \cellcolor{Alto} & \cellcolor{BlazeOrange} \\
\hline
\textit{Backdoor} & \cellcolor{SelectiveYellow} & \cellcolor{VeniceBlue} & \cellcolor{BlazeOrange} & \cellcolor{BlazeOrange} & \cellcolor{VeniceBlue} & \cellcolor{Alto} \\
\hline
\textit{Sinkhole} & \cellcolor{Alto} & \cellcolor{Alto} & \cellcolor{Alto} & \cellcolor{Alto} & \cellcolor{BlazeOrange} & \cellcolor{Alto} \\
\hline
Sybil & \cellcolor{SelectiveYellow} & \cellcolor{Alto} & \cellcolor{Alto} & \cellcolor{SelectiveYellow} & \cellcolor{Alto} & \cellcolor{BlazeOrange} \\
\hline
Masquerade Attack & \cellcolor{BlazeOrange} & \cellcolor{Alto} & \cellcolor{Alto} & \cellcolor{BlazeOrange} & \cellcolor{Alto} & \cellcolor{Alto} \\
\hline
Skimming & \cellcolor{SelectiveYellow} & \cellcolor{VeniceBlue} & \cellcolor{SelectiveYellow} & \cellcolor{BlazeOrange} & \cellcolor{Alto} & \cellcolor{Alto} \\
\hline
Insider & \cellcolor{Alto} & \cellcolor{SelectiveYellow} & \cellcolor{Alto} & \cellcolor{SelectiveYellow} & \cellcolor{SelectiveYellow} & \cellcolor{Alto} \\
\hline
Node Replication & \cellcolor{Alto} & \cellcolor{VeniceBlue} & \cellcolor{Alto} & \cellcolor{SelectiveYellow} & \cellcolor{BlazeOrange} & \cellcolor{Alto} \\
\hline
Wormhole & \cellcolor{Alto} & \cellcolor{BlazeOrange} & \cellcolor{BlazeOrange} & \cellcolor{BlazeOrange} & \cellcolor{VeniceBlue} & \cellcolor{Alto} \\
\hline
Hijacking & \cellcolor{BlazeOrange} & \cellcolor{BlazeOrange} & \cellcolor{Alto} & \cellcolor{BlazeOrange} & \cellcolor{Alto} & \cellcolor{SelectiveYellow} \\
\hline
\textit{Hello} Flooding & \cellcolor{Alto} & \cellcolor{Alto} & \cellcolor{Alto} & \cellcolor{Alto} & \cellcolor{BlazeOrange} & \cellcolor{Alto} \\
\hline
Desynchronization & \cellcolor{Alto} & \cellcolor{BlazeOrange} & \cellcolor{Alto} & \cellcolor{Alto} & \cellcolor{BlazeOrange} & \cellcolor{Alto} \\
\hline
\textit{Blackhole/Grayhole} & \cellcolor{Alto} & \cellcolor{Alto} & \cellcolor{Alto} & \cellcolor{Alto} & \cellcolor{BlazeOrange} & \cellcolor{Alto} \\
\hline
Linkability Attack & \cellcolor{BlazeOrange} & \cellcolor{Alto} & \cellcolor{BlazeOrange} & \cellcolor{BlazeOrange} & \cellcolor{Alto} & \cellcolor{Alto} \\
\hline
\end{tabular}
\end{table}

For each of the described attacks, we performed the STRIDE classification together with criticality assessment using the CVSS, according to Marisetty's example~\cite{arm19}, as seen in Table \ref{tab:stridecvss}. The colors represent the criticality levels according to the CVSS score, as follows: blue is critical (CVSS 9-10), orange is high (CVSS 7-8.9), yellow is medium (CVSS 4-6.9), and gray represents “not applicable” (N/A).

A brief explanation for each classification in Table \ref{tab:stridecvss} is presented below.

\begin{itemize}
    \item \textbf{Eavesdropping} - Threat category \textbf{I} (Information Disclosure), as the focus is the disclosure of data during transmission, and a \textbf{high} criticality due to the direct impact on confidentiality.
    \item \textbf{Brute-Force} - Threat category \textbf{S} (Spoofing), as it attempts to gain access with a false identity by guessing credentials, and a \textbf{medium} criticality.
    \item \textbf{Distributed Denial of Service (DDoS)} - Threat category \textbf{D} (Denial of Service), since the attack's focus is to affect system availability through overload, and a \textbf{high} criticality.
    \item \textbf{Jamming} - Threat categories \textbf{I} and \textbf{D}, given that it can cause information disclosure (I) and affects the availability (D) of the communication channel, and a \textbf{high} criticality for both categories.
    \item \textbf{Sleep Deprivation} - Threat category \textbf{D}, as the attack aims to exhaust the device's battery resource, impairing its availability, and a \textbf{high} criticality for this category.
    \item \textbf{Physical capture of IoT devices} - Threat categories \textbf{I} and \textbf{D}, as an adversary can get information from the captured device (I) or can make it unavailable (D), and \textbf{medium} criticality for both categories.
    \item \textbf{Phishing} - Threat categories \textbf{S, R} (Repudiation), and \textbf{I}, as it encompasses spoofing (S), repudiation (R), and disclosure (I), with the three categories classified with a \textbf{medium} criticality.
    \item \textbf{Spoofing} - Threat categories \textbf{S, T} (Tampering), \textbf{I}, and \textbf{D}, because it involves \textbf{S}, with a \textbf{high} CVSS, through the falsification of IPs, \textbf{T}, with a \textbf{critical} CVSS, because the adversary can gain access and modify the transmitted data, \textbf{I}, with a \textbf{high} CVSS, because the adversary gains access to data and can disclose them, and \textbf{D}, with a \textbf{critical} CVSS, because the adversary can disrupt communication availability.
    \item \textbf{Tampering} - Threat categories \textbf{S, T, R}, and \textbf{I}, given that the focus is data tampering (T), with \textbf{critical} CVSS, but it may also involve spoofing (S) to gain access, with \textbf{high} CVSS, and allows for data repudiation (R), with \textbf{medium} CVSS, and information disclosure (I), with \textbf{high} CVSS.
    \item \textbf{Replay} - Threat categories \textbf{S, R}, and \textbf{I}, as the attack involves retransmitting valid messages repeatedly, obtaining \textbf{high} criticality for all categories.
    \item \textbf{De-authentication} - Threat category \textbf{D} with a \textbf{high} criticality, as the focus is to disconnect legitimate users, affecting availability.
    \item \textbf{Resource Exhaustion} - Threat category \textbf{D} with a \textbf{high} criticality as the attack occurs by depleting memory or CPU.
    \item \textbf{Blueborne} - Threat categories \textbf{T, I,} and \textbf{D}, given that the attack can allow tampering with (T) data or denying (D) access, both with \textbf{high} criticality, and information disclosure (I) with \textbf{medium} criticality.
    \item \textbf{\textit{Man-In-The-Middle} (MITM)} - Threat categories \textbf{S, T, R, I,} and \textbf{E} (Elevation of Privilege), because the adversary intercepts communication, allowing spoofing (S), repudiation (R), and elevation of privilege (E) with \textbf{medium} CVSS, and tampering (T) and disclosure (I) \textbf{high} CVSS.
    \item \textbf{Identity Theft} - Threat categories \textbf{S, T, R, I,} and \textbf{E}, as it is focused on spoofing (S), data tampering (T), information disclosure (I), and elevation of privilege (E), with \textbf{high} CVSS, while also allowing for repudiation (R), with \textbf{medium} score.
    \item \textbf{\textit{Backdoor}} - Threat categories \textbf{S, T, R, I,} and \textbf{D}, with a \textbf{critical} score for T and D, a \textbf{high} score for R and I, and a \textbf{medium} score for S.
    \item \textbf{\textit{Sinkhole}} - Threat category \textbf{D} with a \textbf{high} criticality, given that the malicious node attracts and discards traffic, directly affecting service availability.
    \item \textbf{Sybil} - Threat categories \textbf{S, I,} and \textbf{E}, with a \textbf{medium} score to S and I, and a \textbf{high} score to E, due to the creation of multiple false identities (S) to cause information disclosure (I) and the possibility of gaining elevation of privilege (E).
    \item \textbf{Masquerade Attack} - Threat categories \textbf{S} and \textbf{I} with \textbf{high} score for both, as it is focused on spoofing (S) and information disclosure (I) by impersonating another entity.
    \item \textbf{Skimming} - Threat categories \textbf{S, T, R,} and \textbf{I}, with a \textbf{medium} score for spoofing (S) and repudiation (R), a \textbf{high} score for information disclosure (I), and a \textbf{critical} score for data tampering (T).
    \item \textbf{Insider} - Threat categories \textbf{T, I,} and \textbf{D}, as the internal access allows for data tampering (T), information disclosure (I), and denial (D) of service, both with a \textbf{medium} CVSS score.
    \item \textbf{Node Replication} - Threat categories \textbf{T, I,} and \textbf{D}, with a \textbf{critical} CVSS score for T since the attack can lead to tamper with the device's data, a \textbf{high} score for D because it can deny the service, and a \textbf{medium} score for I as it can disclose information.
    \item \textbf{Wormhole} - Threat categories \textbf{T, R, I,} and \textbf{D}, involving the creation of a malicious tunnel that allows for data tampering (T), information disclosure (I), repudiation (R), both with a \textbf{high} score, and denial of service (D) with \textbf{critical} severity.
    \item \textbf{Hijacking} - Threat categories \textbf{S, T, I,} and \textbf{E}, as it is focused on spoofing (S) and elevation of privilege (E) by taking over a session, allowing for data tampering (T) and information disclosure (I), with \textbf{high} severity for S, T, and I, and \textbf{medium} severity for E.
    \item \textbf{Hello Flooding} - Threat category \textbf{D} with a \textbf{high} criticality, as the focus is to broadcast high-power signals to overwhelm and exhaust network resources, directly affecting availability.
    \item \textbf{Desynchronization} - Threat categories \textbf{T} and \textbf{D} with \textbf{high} criticality for both, given that the adversary tampers (T) with communication parameters to disrupt synchronization and deny (D) service to legitimate nodes.
    \item \textbf{\textit{Blackhole/Grayhole}} - Threat category \textbf{D} with a \textbf{high} criticality, given that the malicious node silently discards network packets to perform a denial (D) of service.
    \item \textbf{Linkability Attack} - Threat categories \textbf{S, R,} and \textbf{I},  with all categories presenting \textbf{high} CVSS scores, as this attack focuses on spoofing (S), repudiation (R) and information disclosure (I).
\end{itemize}

\subsection{Distribution according to STRIDE categories}
\begin{figure}[thb]
    \centering
    \includegraphics[width=\columnwidth]{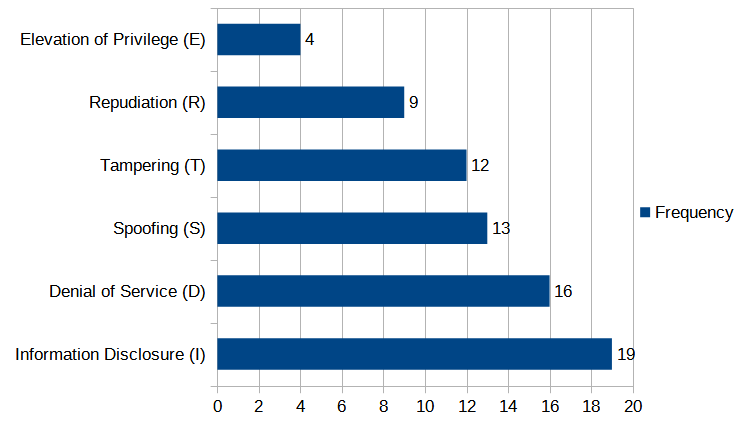}
    \caption{Frequency of attacks per STRIDE category.}
    \label{fig:stride}
\end{figure}

Mapping attacks to the STRIDE categories provides insight into the security properties most frequently violated in IoT environments. As seen in Fig. \ref{fig:stride}, Information Disclosure (I) and Denial of Service (D) are the two most common impacts, identified 19 and 16 times, respectively. The high frequency of Information Disclosure highlights the pervasive confidentiality/privacy risks in IoT, where attacks such as eavesdropping and linkability exploit weak encryption and the lack of secure transmission channels. Similarly, the prevalence of Denial of Service underscores the inherent fragility of resource-constrained devices, which are highly susceptible to battery exhaustion and traffic-based congestion.

Spoofing (S) and Tampering (T) also represent significant portions of the threat landscape, appearing in 13 and 12 attack classifications, respectively. These categories are central to identity-based and integrity-based threats, such as man-in-the-middle (MITM), identity theft, and masquerade attacks, which leverage flaws in authentication and data-handling processes. While Elevation of Privilege (E) was the least frequent category (4 occurrences), it represents the final stage of sophisticated attack chains, such as backdoor or hijacking, where an adversary achieves total control over the vulnerable system. This multi-dimensional view confirms that IoT security cannot rely on a single defensive layer, as attackers exploit diverse technical entry points across the architecture.

\subsection{Distribution according to CVSS severity}
\begin{figure}[t]
    \centering
    \includegraphics[width=\columnwidth]{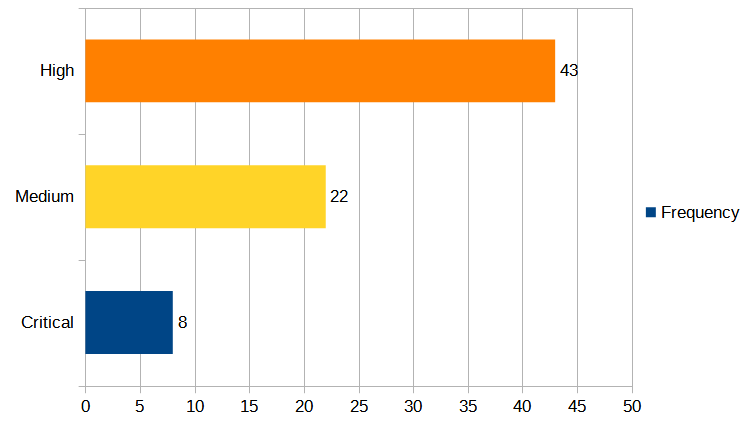}
    \caption{Frequency of attacks per CVSS severity.}
    \label{fig:cvss}
\end{figure}

The quantitative assessment using the CVSS reveals a high-risk profile for the majority of the surveyed attacks. As illustrated in the severity distribution (Fig. \ref{fig:cvss}), the ``High'' criticality level is the most prevalent, accounting for 43 classifications within the studied attack set. This predominance indicates that most IoT threats, such as DDoS, Jamming, and Hello Flooding, possess the potential to significantly disrupt services or compromise sensitive data, requiring urgent mitigation efforts.

A smaller but critical subset of 8 classifications reached the ``Critical'' tier (CVSS 9.0–10.0), primarily associated with attacks that grant adversaries full control over the target system or allow for unauthorized modification of critical data, such as Backdoor, Node Replication, and Wormhole exploits. Conversely, the 22 classifications in the ``Medium'' category often correspond to attacks requiring physical proximity or human interaction, such as Physical Capture, Skimming, and Phishing. This distribution underscores the need for a prioritized defense-in-depth strategy, in which critical architectural vulnerabilities are addressed first, followed by operational and physical security controls.

%% file: vulnattacks.tex
\section{Vulnerability Classification and Attack Mapping}
\label{sec:vulnattacks}

One way to systematically manage and prioritize security efforts is adopting a structured classification scheme for vulnerabilities. In this sense, the following classification scheme categorizes the vulnerabilities into five distinct classes, according to their origin and nature:

\begin{figure*}[!tb]
    \centering
    \includegraphics[width=0.7\textwidth]{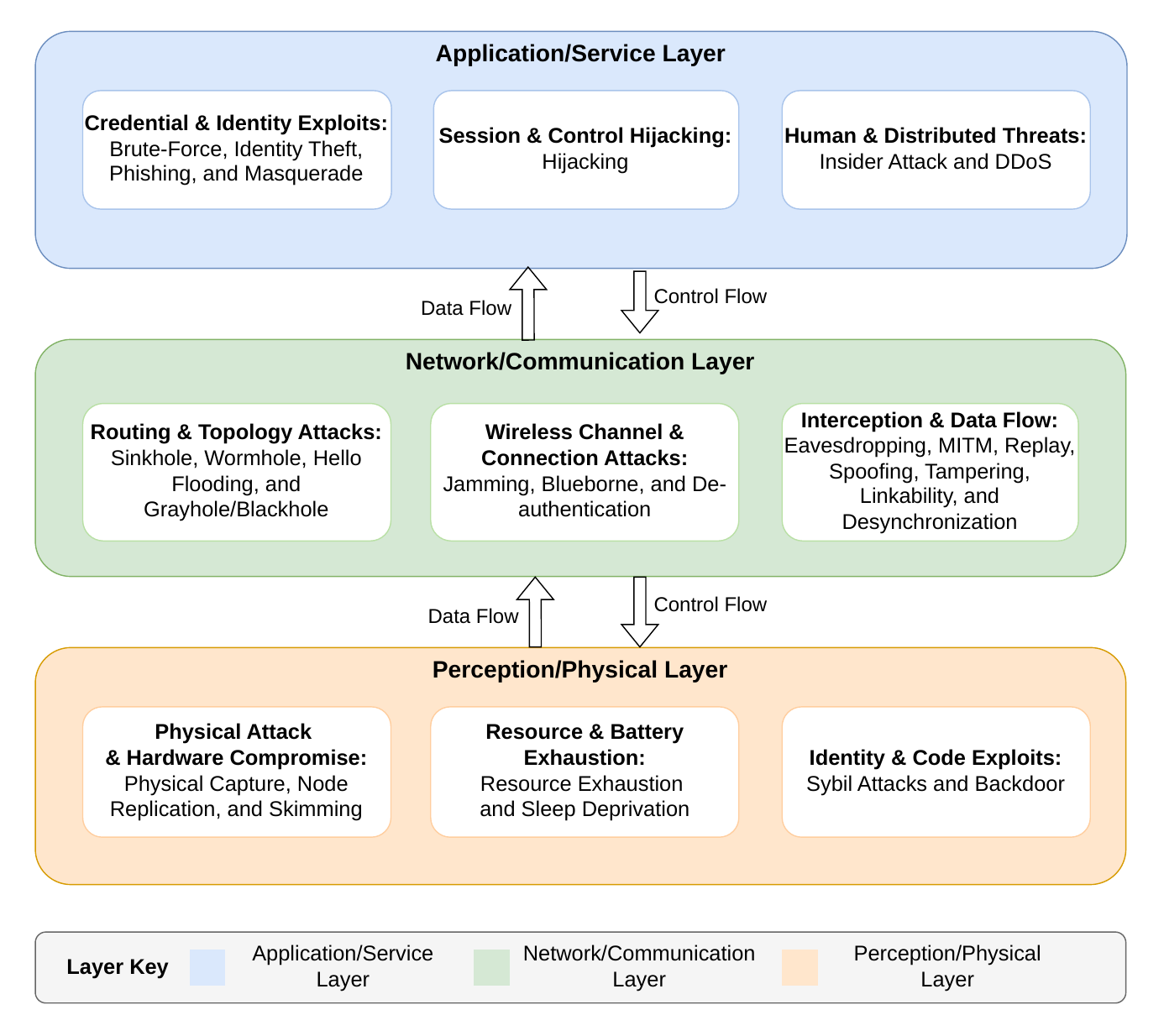}
    \caption{Attacks according to the IoT architectural layers.}
    \label{fig:arch_attacks}
\end{figure*}

\begin{itemize}
    \item Process - Weaknesses occurring in common security protocols and functions, such as authentication, attestation, access control, and data handling workflows.
    \item Code - Flaws directly related to the development process, encompassing insecure programming practices, buffer overflows, and the use of vulnerable cryptographic algorithms or outdated protocols within the source code.
    \item Communication - Vulnerabilities inherent to network protocols, data transmission mechanisms, and the wireless channel itself, often related to the lack of encryption or secure pairing procedures.
    \item Operation - Security issues stemming from human factors or environmental configurations, including the use of default/weak credentials, outdated software/firmware, and misconfiguration of operating systems or security settings.
    \item Device - Weaknesses tied to the intrinsic physical and computational characteristics of the IoT hardware, such as resource constraints, weak physical security, and insecure communication interfaces.
\end{itemize}

\begin{table*}[!tb]
\centering
\scriptsize
\setlength{\tabcolsep}{3pt}
\renewcommand{\arraystretch}{0.9}
\caption{Mapping of common IoT attacks to vulnerability classes.}
\label{tab:attack_mapping}
\begin{adjustbox}{width=\textwidth}
\begin{tabular}{@{}p{3cm} p{3cm} p{3.5cm} p{7.5cm}@{}}
\toprule
\textbf{Attack Type} & \textbf{Primary Focus} & \textbf{Vulnerability Classes} & \textbf{Justification} \\ \midrule

Eavesdropping & Data interception & Communication, Process & Exploits lack of encryption and secure transmission channels (communication). Can leverage RPCs within the process for full device control. \\ \addlinespace

Brute-Force & Weak credentials & Operation, Process & Targets the use of default or easily guessable passwords (a flaw in operation). Successful guessing bypasses the authentication process. \\ \addlinespace

DDoS & Traffic overload & Communication, Operation & Overwhelms systems by exploiting network capacity (communication). Often launched via botnets formed by devices compromised due to poor operation. \\ \addlinespace

Jamming & Signal interference & Communication, Device & Exploits the susceptibility of wireless communication technologies to interference. Drains batteries and forces retransmissions. \\ \addlinespace

Sleep Deprivation & Battery exhaustion & Device, Process & Exploits restricted power device resources. Disrupts the normal sleep-wake cycle by causing constant activity. \\ \addlinespace

Physical Capture & Hardware access & Device, Operation & Exploits weak device physical security. Gaining physical possession allows an attacker to impersonate a legitimate user. \\ \addlinespace

Phishing & Social engineering & Operation, Process & Exploits human behavior to illicitly acquire sensitive credentials. Successful theft bypasses the authentication process. \\ \addlinespace

Spoofing & Identity forgery & Communication, Process & Attacker disguises the identity. Methods like IP or ARP spoofing directly manipulate communication protocols. \\ \addlinespace

Tampering & Data modification & Communication, Code & Involves intercepting communication to modify device codes or data. Exploits flaws in code integrity checks. \\ \addlinespace

Replay Attack & Message re-transmission & Process, Communication & Relies on intercepting authentication signals and retransmitting them, exploiting a failure in session validation. \\ \addlinespace

De-authentication & Forced disconnection & Communication & Involves deliberately sending control packets to disconnect targeted devices, exploiting management stability. \\ \addlinespace

Resource Exhaustion & Resource depletion & Device & Overwhelms key device resources such as memory, CPU, or battery. \\ \addlinespace

Blueborne & Bluetooth exploitation & Communication, Device & Exploits vulnerabilities in Bluetooth communication protocols. Affects a high percentage of Bluetooth-equipped devices. \\ \addlinespace

Man-In-The-Middle & Intercepting comm. & Communication, Process & Intercepts and potentially modifies data. Exploits the lack of mutual authentication process. \\ \addlinespace

Identity Theft & User credential theft & Operation, Process & Achieved by exploiting default/weak credentials or social engineering, bypassing the authentication process. \\ \addlinespace

Backdoor & Security bypass & Code, Process & Involves installing malicious code to bypass security mechanisms or evade the authentication process. \\ \addlinespace

Sinkhole & Routing manipulation & Communication & Deploys a deceptive node to broadcast false routing information, manipulating the flow of network communication. \\ \addlinespace

Sybil & False identities & Device, Process & Adversary creates multiple false identities, exploiting the lack of robust authentication due to resource constraints. \\ \addlinespace

Masquerade & Impersonation & Process, Operation & Uses forged/stolen credentials to impersonate users. Careless user behavior can facilitate the attack. \\ \addlinespace

Skimming & Credential cloning & Device, Operation & Requires a physical skimmer device near a reader to steal credentials used in operation processes. \\ \addlinespace

Insider Attack & Trusted actor damage & Operation, Code & An internal actor with legitimate access intentionally or unintentionally alters configuration or code. \\ \addlinespace

Node Replication & Node cloning & Device, Process & Attacker captures a node and creates identical copies, disrupting the node identification process. \\ \addlinespace

Wormhole & Covert tunnel & Communication, Process & Creates a secret communication tunnel between nodes to intercept data, bypassing standard routing. \\ \addlinespace

Hijacking & Control seizure & Process, Operation & Gains unauthorized control. Often relies on exploiting weak passwords or session IDs. \\ \addlinespace

Hello Flooding & Routing congestion & Communication, Process & Floods nodes with HELLO messages, exploiting trust in neighbor discovery to cause congestion. \\ \addlinespace

Desynchronization & Authentication disruption & Process, Communication & Disrupts the authentication process by breaking synchronization between client and access point. \\ \addlinespace

Grayhole/Blackhole & Selective forwarding & Communication, Process & Nodes discard packets, disrupting the flow of communication. Exploits flaws in routing and packet forwarding. \\ \addlinespace

Linkability Attack & Correlation of identity & Communication & Adversary correlates activities to the same source by exploiting message patterns and device identifiers. \\

\bottomrule
\end{tabular}%
\end{adjustbox}
\end{table*}

Classifying IoT vulnerabilities into structured classes is fundamental to a proactive defense, as it enables the implementation of layered security measures tailored to the specific nature of each threat. By identifying whether a flaw resides in the Process, Code, Communication, Operation, or Device category, organizations can assign precise mitigation responsibilities, such as requesting manufacturers to fix hardware-level device flaws while IT teams manage operational configurations and monitoring. Mapping these vulnerability classes directly to known attacks is of paramount importance, as it allows security professionals to visualize the specific entry points and technical weaknesses exploited by the attacks. This coordinated approach ensures that security hardening is not just a reactive measure, but a strategic alignment where operational processes and technical fixes work together to create a resilient IoT environment.

Based on that classification, we conducted an assessment of the described IoT attacks to establish a direct correlation between them and the root cause vulnerability classes they exploit. The resulting mapping, presented in Table \ref{tab:attack_mapping}, defines the scope of each attack vector, linking observed attack techniques to the underlying security weakness categories.

Besides this mapping, considering the vulnerability classes and attacks, we can also categorize the attacks according to each of the three layers from the basic IoT architecture (i.e., perception/physical layer, network/communication layer, and application/service layer), as shown in Figure \ref{fig:arch_attacks}. Thus, in the following subsections, we list the attacks separated by each layer.

\subsection{Perception/Physical Layer} The attacks grouped here exploit physical vulnerabilities, hardware constraints (e.g., battery and memory), and authentication weaknesses at the node level. They focus on the following vulnerability classes: Device and Operation.
    
\begin{itemize}
    \item Physical attack and hardware compromise
    \begin{itemize}
        \item \textbf{Physical capture} of IoT devices: gaining physical possession of hardware to extract data or modify components.
        \item \textbf{Node replication}: capturing a legitimate node to create and deploy unauthorized, identical clones.
        \item \textbf{Skimming}: attaching malicious hardware near a legitimate reader to steal credentials during operation.
    \end{itemize}
    
    \item Resource and battery exhaustion
    \begin{itemize}
        \item \textbf{Resource exhaustion}: overwhelming a node's CPU or memory to cause service disruption.
        \item \textbf{Sleep deprivation}: orchestrating interactions that prevent a node from entering power-saving sleep modes, rapidly draining its battery.
    \end{itemize}
    
    \item Identity and code exploits
    \begin{itemize}
        \item \textbf{Sybil attack}: creating multiple fake identities at the node level to undermine voting or reputation systems.
        \item \textbf{Backdoor}: injecting malicious code into the device's firmware or software to bypass local security controls.
    \end{itemize}
\end{itemize}

\subsection{Network/Communication Layer}
This is the densest layer, positioning attacks that affect data in transit and network topology, as this layer involves data transmission, routing (e.g., RPL), and wireless communication channels (e.g., Wi-Fi, Bluetooth, 5G). The attacks in this layer has the focus on the following vulnerability classes: Communication and Process.
\begin{itemize}
    \item Routing and topology attacks
    \begin{itemize}
        \item \textbf{Sinkhole}: manipulating routing protocols to attract all traffic through a malicious node.
        \item \textbf{Wormhole}: establishing a secret, low-latency tunnel between two distant nodes to misdirect traffic.
        \item \textbf{Hello Flooding}: congesting the network by broadcasting ``HELLO'' messages, exploiting neighbor discovery processes.
        \item \textbf{Grayhole/Blackhole}: selectively (Grayhole) or completely (Blackhole) dropping packets to disrupt end-to-end communication.
    \end{itemize}
    \item Wireless channel and connection attacks
    \begin{itemize}
        \item \textbf{Jamming}: generating radio frequency interference to disrupt legitimate wireless signals.
        \item \textbf{Blueborne}: exploiting vulnerabilities in Bluetooth implementations to gain unauthorized access or execute code.
        \item \textbf{De-authentication attack}: forcibly disconnecting legitimate nodes from a wireless network.
    \end{itemize}
    \item Interception and data flow attacks:
    \begin{itemize}
        \item \textbf{Eavesdropping}: passively intercepting unencrypted data transmitted over the wireless channel.
        \item \textbf{Man-in-the-Middle (MITM)}: actively intercepting, and potentially modifying, the data flow between two communicating parties.
        \item \textbf{Replay attack}: capturing valid messages and retransmitting them later to produce an unauthorized effect.
        \item \textbf{Spoofing}: forging IP or ARP addresses to impersonate legitimate devices and manipulate network protocols.
        \item \textbf{Tampering}: maliciously modifying data packets while they are in transit across the network.
        \item \textbf{Linkability attack}: correlating different identities, messages, or routes to track a user's activity.
        \item \textbf{Desynchronization}: breaking the synchronization between a client and an access point to disrupt active sessions.
    \end{itemize}
\end{itemize}

\subsection{Application/Service Layer}
The attacks here exploit session management, credentials, social engineering, and human-centric configurations, as this topmost layer involves servers, dashboards, APIs, and direct interaction with human users or control systems. The attacks listed below focus on the following vulnerability classe: Process and Operation.

\begin{itemize}
    \item Credential and identity exploits:
    \begin{itemize}
        \item \textbf{Brute-force attack}: systematically attempting to guess user credentials or cryptographic keys.
        \item \textbf{Identity theft}: stealing legitimate credentials through malware, social engineering, or by exploiting authentication flaws.
        \item \textbf{Phishing}: using social engineering to trick users into revealing sensitive information.
        \item \textbf{Masquerade attack}: using stolen or forged credentials to gain unauthorized access to system resources.
    \end{itemize}
    
    \item Session and control hijacking:
    \begin{itemize}
        \item \textbf{Hijacking}: taking over legitimate user sessions or gaining unauthorized control of IoT management systems.
    \end{itemize}
    \item Human and distributed threats:
    \begin{itemize}
        \item \textbf{Insider attack}: malicious actions performed by a trusted entity with legitimate access to the system.
        \item \textbf{Distributed Denial of Service (DDoS)}: overwhelming application-layer services or main gateways with a massive volume of distributed traffic, often leveraging botnets of compromised devices.
    \end{itemize}
\end{itemize}

\subsection{Security Planning and Monitoring}
This taxonomic approach serves as a critical, actionable tool for organizations looking to design resilient IoT systems and establish robust operational security. By transitioning from a generalized view of ``cyber threats'' to a structured, context-aware framework, stakeholders can significantly improve both their proactive security planning and reactive security monitoring.

\subsubsection{Proactive Security Planning and Design}

The primary utility of these mappings in the planning phase lies in enabling a methodical, defense-in-depth approach tailored to the unique characteristics of an IoT deployment.

First, by referencing the layered taxonomy (illustrated in Fig. \ref{fig:arch_attacks}), security architects can identify which components face specific classes of threats. For instance, knowing that the Perception/Physical Layer is  susceptible to physical exploits and resource exhaustion allows designers to incorporate hardware-based security modules (HSMs), tamper-responsive circuitry, and robust power-management techniques that are highly relevant to that specific context. Conversely, planners focused on the Application/Service Layer can prioritize secure API development, multi-factor authentication, and sophisticated session management without over-engineering individual sensors.

Furthermore, the mapping between vulnerabilities (e.g., Code, Operation, Communication) and specific attacks allows for targeted control implementation. A planning team can cross-reference high-CVSS scoring attacks like ``Wormhole'' or ``Sinkhole'' (Network Layer) against the ``Communication'' vulnerability class. This direct correlation provides a clear mandate to prioritize robust, authenticated routing protocols rather than allocating resources to generic perimeter defenses that cannot stop these internal, topology-based attacks. This approach ensures that security investments are data-driven and effectively address the technical entry points most likely to be exploited.

\subsubsection{Reactive Security Monitoring and Incident Response}

Once an IoT system is operational, these structured taxonomies continue to provide significant value by enhancing the accuracy of security monitoring and accelerating incident response times.

In traditional IT networks, security operation centers (SOCs) are often overwhelmed by a high volume of generic alerts. However, in an IoT environment mapped by layers, detection systems (like IDS/IPS) can be configured with context-aware rules. For example, anomaly detection rules at the Network/Communication Layer can be specifically tuned to identify patterns characteristic of routing manipulation attacks (e.g., Sinkhole and Blackhole), such as a sudden shift in traffic paths or excessive packet drops from a single node. Monitoring systems do not need to look for generic malware; they are configured to look for the types of attacks known to manifest at that specific architectural interface, leading to significantly fewer false positives and a higher detection rate for sophisticated threats.

Crucially, the attack-layer mapping drastically accelerates the triage and containment phases of incident response. When a security alert fires, the responder immediately understands the technical domain of the threat and the likely entry point. If the monitoring system flags a potential ``Elevation of Privilege'' attempt, the operators can proceed — based on the taxonomy in Fig. \ref{fig:arch_attacks} — to immediately isolate components at the Application/Service Layer, prioritize checking Application logs, and audit user sessions. They do not lose critical time inspecting network packet captures for topology changes. Similarly, an alert concerning ``Sleep Deprivation'' or ``Resource Exhaustion'' immediately redirects investigators to the Perception/Physical Layer, prompting physical inspection of hardware or memory dump analysis of individual nodes. By reducing ambiguity, the taxonomic mapping ensures that incident response teams can contain threats faster and trace root causes with greater precision.

%% file: vertattacks.tex
\section{Main Attacks per IoT Vertical}
\label{sec:vert}
The heterogeneity of the IoT ecosystem dictates that not all attacks are equally probable or impactful across different application domains. Device placement, communication protocols, and the nature of the processed data significantly alter the threat scenario. This section maps the most prevalent attacks from our taxonomy to the four major IoT verticals, highlighting the specific vulnerabilities that adversaries are most likely to exploit in each scenario.

It is important to note that the attacks mapped in this section are not mutually exclusive to a single vertical. Since diverse IoT applications share a common underlying architecture -- comprising perception, network, and application layers -- and rely on similar communication protocols, significant threat intersections exist. For instance, while Distributed Denial of Service (DDoS) is a common threat in Smart Cities due to the massive scale of exploitable devices, it is equally devastating in Industry 4.0, where it can halt critical production lines. Similarly, Eavesdropping poses severe privacy risks in Digital Health, but also facilitates industrial espionage in IIoT environments. Therefore, the mapping provided in the following subsections highlights the threats that are most prevalent or have the most critical context-specific impact within each domain, acknowledging that the broader threat landscape remains highly interconnected.

\subsection{Smart Cities}
In urban environments, the security challenge stems primarily from the massive scale and public accessibility of the deployed infrastructure. Devices such as smart streetlights and connected cameras are often physically exposed, making them highly susceptible to tampering and hardware compromise. Furthermore, the sheer volume of interconnected nodes creates an ideal breeding ground for large-scale botnet recruitment. As demonstrated by historical incidents, such as the Mirai botnet, compromised urban infrastructure can be leveraged to launch devastating Distributed Denial of Service (DDoS) attacks, crippling critical city services and communication networks. Table \ref{tab:smart_cities_attacks} presents the four most common attacks to this vertical.

\begin{table}[!ht]
\centering
\caption{Prevalent Attacks in Smart Cities}
\label{tab:smart_cities_attacks}
\renewcommand{\arraystretch}{1.2}
\begin{tabular}{|l|p{5.5cm}|}
\hline
\textbf{Attack} & \textbf{Rationale in this Vertical} \\ \hline
\textbf{Physical Capture} & Devices (e.g., smart streetlights, cameras) are in public spaces, making hardware easily accessible to adversaries. \\ \hline
\textbf{DDoS} & The massive scale of urban IoT makes it a perfect breeding ground for botnets to overwhelm city services. \\ \hline
\textbf{Skimming} & Often deployed near public access control systems or smart parking payment terminals to clone credentials. \\ \hline
\textbf{Jamming} & Can disrupt wide-area urban networks, affecting intelligent traffic controls and public safety communications. \\ \hline
\end{tabular}
\end{table}

\subsection{Digital Health (Healthcare)}
The threat model in the Internet of Medical Things (IoMT) is defined by the strict necessity for patient data confidentiality and the continuous availability of life-critical devices. Since many medical applications rely on wearable sensors communicating via personal area networks (such as Bluetooth), they present unique vulnerabilities to proximity-based exploits and data interception. In this domain, cyber threats extend beyond data breaches; attacks that deplete a device's battery (e.g., Sleep Deprivation) or manipulate transmitted health metrics can have direct, severe, and potentially fatal consequences for the patients. Table \ref{tab:health_attacks} presents the four most common attacks to this vertical.

\begin{table}[!th]
\centering
\caption{Prevalent Attacks in Digital Health (IoMT)}
\label{tab:health_attacks}
\renewcommand{\arraystretch}{1.2}
\begin{tabular}{|l|p{5.5cm}|}
\hline
\textbf{Attack} & \textbf{Rationale in this Vertical} \\ \hline
\textbf{Eavesdropping} & High risk of interception of highly sensitive personal health data transmitted by wearables. \\ \hline
\textbf{Blueborne} & Wearable medical devices rely heavily on Bluetooth, making them highly susceptible to this protocol-specific exploit. \\ \hline
\textbf{Sleep Deprivation} & Draining the battery of life-sustaining devices can have direct and critical physical consequences. \\ \hline
\textbf{Tampering} & Malicious modification of medical parameters in transit can lead to incorrect and dangerous clinical decisions. \\ \hline
\end{tabular}
\end{table}

\subsection{Industry 4.0 (Industrial IoT)}
Security in the Industrial Internet of Things (IIoT) is fundamentally tied to the protection of physical assets and the continuous operation of manufacturing lines. The convergence of legacy industrial control systems, which originally lacked built-in security, with modern internet-facing networks has significantly expanded the attack surface. The network architecture often involves legacy industrial systems newly exposed to the Internet, making them vulnerable to malicious injections, industrial espionage, and sabotage through unauthorized commands. In these environments, adversaries focus on establishing persistent remote access or manipulating operational commands. Even temporary disruptions, unauthorized data modification, or resource exhaustion at the IIoT gateway level can lead to severe economic losses and physical damage to the infrastructure. Table \ref{tab:industry_attacks} presents the four most common attacks to this vertical. 

\begin{table}[!th]
\centering
\caption{Prevalent Attacks in Industry 4.0 (IIoT)}
\label{tab:industry_attacks}
\renewcommand{\arraystretch}{1.2}
\setlength{\tabcolsep}{4pt}
\begin{tabular}{|l|>{\raggedright\arraybackslash}p{5.5cm}|}
\hline
\textbf{Attack} & \textbf{Rationale in this Vertical} \\ \hline
\textbf{Backdoor} & Attackers inject malicious code to bypass security and establish persistent remote control over industrial gateways. \\ \hline
\textbf{Insider Attack} & Disgruntled or compromised employees can intentionally or accidentally alter critical industrial configurations. \\ \hline
\textbf{Replay Attack} & Capturing and resending valid authentication signals or operational commands can disrupt automated production lines. \\ \hline
\textbf{Resource Exhaustion} & Overwhelming IIoT gateways and PLCs can halt production asset management systems completely. \\ \hline
\end{tabular}
\end{table}

\subsection{Precision Agriculture (Agriculture 4.0)}
Agricultural IoT applications typically rely on highly distributed and unattended Wireless Sensor Networks (WSNs) spread across vast open fields. From a security perspective, this vertical's primary weakness is the absolute lack of physical boundaries combined with the severe energy constraints of the individual sensing nodes. Because these nodes are physically unprotected and operate on severe energy constraints, they are highly vulnerable to physical cloning and routing manipulation attacks that exploit multi-hop topologies. Attackers can easily access the hardware to capture devices, clone them, and inject false environmental data into the system. Additionally, the multi-hop routing topologies used to transmit data over long distances make these networks particularly vulnerable to complex routing manipulation and topology-based attacks. Table \ref{tab:agriculture_attacks} presents the four most common attacks to this vertical. 

\begin{table}[!tp]
\centering
\caption{Prevalent Attacks in Precision Agriculture}
\label{tab:agriculture_attacks}
\renewcommand{\arraystretch}{1.2}
\setlength{\tabcolsep}{4pt}
\begin{tabular}{|l|>{\raggedright\arraybackslash}p{5.5cm}|}
\hline
\textbf{Attack} & \textbf{Rationale in this Vertical} \\ \hline
\textbf{Node Replication} & Easy physical access in open fields allows attackers to capture and clone legitimate sensors to inject false environmental data. \\ \hline
\textbf{Sinkhole} & Attackers deploy a deceptive node in the wide-area mesh network to silently attract and drop critical agricultural data. \\ \hline
\textbf{Hello Flooding} & Exploits the neighbor discovery process in WSNs, rapidly draining the energy of surrounding remote field sensors. \\ \hline
\textbf{Sybil} & Generating multiple false identities in distributed field sensors to manipulate voting or data aggregation algorithms. \\ \hline
\end{tabular}
\end{table}

%% file: FutureDirections.tex
\section{Challenges and Opportunities}
\label{sec:FutureDirections}



Securing the IoT is an ongoing challenge due to scale, heterogeneity, and limited device resources. Billions of devices operate in untrusted environments with minimal computational and energy budgets, making them attractive targets. While standards and security frameworks provide a foundation, several gaps remain, and new opportunities are emerging. Table~\ref{tab:summary} provides an overview of the main challenges, research gaps, and corresponding opportunities for IoT security, including IoT-NDN as a future direction.

\begin{table}[ht]
\centering
\caption{IoT security: challenges, research gaps, and opportunities}
\label{tab:summary}
\begin{tabular}{|>{\raggedright\arraybackslash}p{3.7cm}|>{\raggedright\arraybackslash}p{4.2cm}|}
\hline
\textbf{Challenges / Gaps} & \textbf{Opportunities / Future directions} \\ \hline
Constrained, heterogeneous devices & Edge offloading, lightweight cryptography \\ \hline
Supply-chain vulnerabilities & Secure boot, authenticated updates, enforced baselines \\ \hline
Identity and access at scale & Zero Trust, namespace-driven access control (NAC) \\ \hline
Botnets, large-scale DDoS & ML/DL anomaly detection \\ \hline
Privacy leakage & Privacy-preserving ML, confidential computing \\ \hline
Post-Quantum Criptography (PQC) adoption difficulty & Hybrid crypto, lightweight PQC \\ \hline
Blockchain scalability & Hybrid or lightweight ledger solutions \\ \hline
Lack of monitoring tools & Open-source frameworks for IoT attack detection \\ \hline
NDN-specific threats (e.g., cache poisoning) & Trust schemas, adaptive forwarding, secure caching \\ \hline
Policy fragmentation & Harmonized international IoT security standards \\ \hline
\end{tabular}
\end{table}

\subsection{Key Challenges and Research Gaps}

\begin{enumerate}
    \item \textbf{Resource constraints and heterogeneity} - IoT devices often cannot support strong cryptography or real-time monitoring. Diverse protocols hinder interoperability and coordinated defenses~\cite{roman2013, sicari2015}.
    
    \item \textbf{Device lifecycle security} - Vulnerabilities may be introduced at any stage. Despite standards such as NISTIR 8259A and ETSI EN 303 645, adoption is inconsistent~\cite{nist8259a, etsi2024}.
    
    \item \textbf{Identity and access control} - Scalable credential management and revocation remain unresolved, especially in dynamic IoT deployments~\cite{alaba2017}.
    
    \item \textbf{Large-scale attacks} - Botnets such as Mirai show how insecure devices can launch devastating DDoS attacks~\cite{mirai}.
    
    \item \textbf{Privacy} - Continuous sensing creates risks of data misuse and inference attacks~\cite{roman2013, sicari2015}.
    
    \item \textbf{Cryptographic agility} - The transition to post-quantum cryptography (PQC) is difficult given constrained hardware~\cite{nistpqc}.
    
    \item \textbf{Monitoring and tooling} - Few standardized tools exist to continuously monitor IoT applications and detect well-known attacks. This is a significant research gap.
    
    \item \textbf{Policy and regulation} - National and regional regulations (e.g., NIST, ETSI) set minimum requirements, but harmonization and global enforcement are lacking.
\end{enumerate}

\subsection{Emerging Opportunities and Innovative Solutions}

\begin{enumerate}
    \item \textbf{Security baselines and policies} - Minimum requirements such as secure updates, unique credentials, and vulnerability disclosure (ETSI EN 303 645, NISTIR 8259A) provide a practical foundation, but stronger enforcement and global alignment are needed~\cite{etsi2024, nist8259a}.
    
    \item \textbf{Zero Trust} - Applying Zero Trust principles (NIST SP 800-207) - continuous device authentication, micro-segmentation, and posture validation - can limit attack propagation~\cite{nistzerotrust}.
    
    \item \textbf{AI-assisted security} - Machine learning and deep learning models can detect anomalies, but most results rely on simulation datasets. Research must prioritize real-world IoT datasets and federated learning approaches to preserve privacy while improving detection accuracy~\cite{mlsurvey, flsurvey}.
    
    \item \textbf{Blockchain critique} - Many proposals suggest blockchain-based IoT security. While blockchains ensure integrity and auditability, scalability, latency, and energy costs limit viability. Lightweight alternatives or hybrid ledgers are needed.
    
    \item \textbf{Edge-assisted security} - Gateways and edge nodes can perform heavy cryptographic operations, anomaly detection, and update orchestration, thus protecting constrained devices.
    
    \item \textbf{Security monitoring tools} - A promising direction is the development of open-source, standardized tools to continuously test IoT applications against known attack patterns.
    
    \item \textbf{Post-Quantum Cryptography (PQC) migration.} Hybrid deployments (classical + PQC) and lightweight schemes are necessary for IoT adoption~\cite{nistpqc}.
\end{enumerate}

\subsection{IoT over Named Data Networking (IoT-NDN)}

The current IoT ecosystem largely relies on IP-based communication, where security mechanisms are typically endpoint-oriented and often retrofitted to constrained devices. In contrast, Named Data Networking (NDN) is a network architecture that shifts the communication model from host-to-host addressing to \emph{data-centric} communication~\cite{zhang2014}. Instead of asking \emph{who} provides the data, applications simply request \emph{what} data are needed. Every NDN data packet must be cryptographically signed, embedding authenticity directly into the network layer. This paradigm aligns naturally with IoT requirements and provides several advantages, as follows.

\begin{itemize}
    \item \textbf{Data-centric security} - In NDN, each data packet is authenticated, ensuring integrity and provenance regardless of the transport path or the device providing it. This reduces reliance on device-to-device trust, which is often weak in IoT~\cite{shang2016}.
    
    \item \textbf{Resilience through caching} - Intermediate nodes can cache authenticated data and serve future requests. This property is particularly beneficial in IoT, where devices may frequently disconnect due to mobility, power saving, or lossy links.
    
    \item \textbf{Fine-grained access control} - Mechanisms such as Name-based Access Control (NAC) tie cryptographic keys to namespaces, allowing scalable and flexible enforcement of security policies for different IoT applications~\cite{zhang2018}.
    
    \item \textbf{Reduced attack surface} - Since applications verify \emph{data} instead of \emph{endpoints}, certain classes of man-in-the-middle or spoofing attacks are inherently harder to execute.
\end{itemize}

Recent work has demonstrated the feasibility of integrating NDN into IoT environments~\cite{hail2022}, showing that data-centric approaches can overcome many of the limitations of IP-based systems. In this work, the authors explored the integration of IoT-NDN with FIWARE middleware \footnote{https://www.fiware.org/}, highlighting how such an approach enables scalable trust management and efficient data dissemination. In another work \cite{hail2019}, the authors proposed a lightweight NDN-based architecture tailored for constrained IoT devices, embedding security into the communication process while reducing the overhead associated with heavy cryptographic protocols. IoT-NDN is also applied to caching strategies and access control mechanisms designed specifically to improve efficiency and enforce fine-grained policies in large-scale deployments \cite{amadeo2014}.

IoT deployments increasingly demand security mechanisms that scale with billions of devices, tolerate intermittent connectivity, and support decentralized trust models. NDN directly addresses these requirements by making \emph{security intrinsic to data} and leveraging in-network caching to improve resilience. Compared to IP networks, which requires additional layers and protocols (e.g., TLS, DTLS, VPNs) to provide similar assurances, NDN embeds security into the architecture itself. This makes IoT-NDN an attractive direction for next-generation secure IoT systems.

Despite its promise, IoT-NDN faces the following open research issues:
\begin{itemize}
    \item \textbf{Interest flooding attacks} - NDN routers maintain per-request state, which can be exploited by flooding interest packets, leading to PIT (Pending Interest Table) exhaustion~\cite{gasti2013};
    \item \textbf{Content poisoning} - Malicious data can be injected into caches, polluting results for other consumers~\cite{ghali2014};
    \item \textbf{Privacy concerns} - Semantic information embedded in names and cache timing can leak user behavior;
    \item \textbf{Key management} - Large-scale IoT deployments demand efficient mechanisms for credential distribution, revocation, and post-quantum readiness.
\end{itemize}

Ongoing research is investigating adaptive forwarding to mitigate Interest Flooding, ranking and verification mechanisms for cache protection, and namespace-based trust schemas for scalable key management. Combining IoT-NDN with machine learning (e.g., anomaly detection in content requests) or edge offloading of signature verification offers further opportunities. Moreover, the integration of lightweight post-quantum cryptographic schemes will be essential for long-term security. Addressing these gaps will be crucial to make IoT-NDN a practical and secure alternative to IP-based IoT systems.

\subsection{Artificial Intelligence for IoT}

The intersection of AI and IoT has become the primary theater for addressing IoT's core security challenges. Recent works have highlighted this trend, where AI-assisted methodologies are used to harden vulnerable entry points, moving beyond signature-based detection toward contextual and behavioral intelligence~\cite{mohamed2025artificial,rafique2024machine}. A recent cybersecurity review further confirms that AI is now being used to strengthen intrusion detection, malware classification, behavioral analysis, and threat intelligence, while also introducing new concerns such as adversarial machine learning and limited explainability~\cite{ogenyi2025securing}.

Security challenges in IoT stem from the inherent limitations of devices (e.g., limited computational resources, heterogeneous protocols, and a lack of standardized update mechanisms). However, IoT vulnerabilities are widespread and can be categorized into distinct classes that define the technical entry points for adversaries. AI has been applied to mitigate each of them.

Traditional process management relies on static rules that are easily bypassed by dynamic social engineering or protocol-level manipulation, creating a significant entry point for adversaries seeking process vulnerabilities. A significant related advancement is the emergence of the Intelligent Policy Agent Network (IPAN)~\cite{segal2026autonomous}. IPAN replaces these static rules with LLM-powered agents that operate as autonomous policy enforcers. Empirically, these frameworks have demonstrated a 94\% improvement in policy compliance rates and a 67\% reduction in unauthorized access incidents.

AI has demonstrated the feasibility of mitigating phishing, brute-force, and other attacks related to process vulnerabilities. A recent work introduced AI systems incorporating Kernel-Based Principal Component Analysis (KPCA) and VGG-16 classification to enable highly accurate biometric-based intrusion detection~\cite{alfahaid2025machine}. Combined with the Deterministic Trust Transfer Protocol (DTTP), these systems achieve 96\% accuracy in ensuring data integrity and identity verification throughout the device lifecycle. By embedding identity within the data flow itself, AI mitigates the risk of session hijacking and masquerade attacks.

AI has also been applied to mitigate code vulnerabilities. Decoding-enhanced BERT with Disentangled Attention (DeBERTa) was used for detecting vulnerabilities in firmware binaries \cite{nandish2025transformer}. The disentangled attention mechanism allows the model to process content and position information separately, enhancing its ability to reason about complex control flows and data structures within the binary. This approach captures deep bidirectional dependencies, enabling the identification of subtle patterns associated with buffer overflows and use-after-free errors. Experimental findings demonstrate that this DeBERTa-based model achieves 97\% accuracy and a 94.6\% F1-score, significantly outperforming conventional embedding techniques.

Still focusing on flaws in the development process, researchers have developed pipelines that can analyze unstructured code and extract contextual meaning~\cite{mane2025automated}. This is particularly effective for "black-box" systems where traditional static analysis fails. These pipelines automate the identification of binary targets and produce comprehensive vulnerability reports, identifying high-risk flaws such as SQL injection (CWE-89) and insecure innerHTML operations (CWE-79) with greater speed and accuracy than human analysts.
Furthermore, AI-driven automation can facilitate the creation of Software Bill of Materials (SBOMs)~\cite{nandish2025transformer}. This inventory allows security teams to track third-party library dependencies and identify components with known CVEs. This can be essential for preventing supply chain attacks, where malicious code is embedded during the manufacturing phase~\cite{nandish2025transformer}.

The use of Graph Neural Networks (GNNs) to model the structural and temporal relationships of IoT networks indicates the opportunity to use AI to mitigate IoT communication vulnerabilities~\cite{alotaibi2025ai}. By modeling the IoT network as a dynamic constrained graph, GNNs can harness belief propagation to verify the legitimacy of neighbor discovery and routing announcements~\cite{alotaibi2025ai}. For example, GNN-driven adaptive routing frameworks have been proposed to optimize communication resilience by predicting future performance metrics and adjusting paths to bypass jammed or compromised nodes~\cite{li2025achieving}. In the context of Software-Defined IoT (SD-IoT), GNNs are integrated with Transformers and Reinforcement Learning agents to create hybrid intrusion detection and mitigation frameworks~\cite{alotaibi2025ai}.

While GNNs handle the structural interaction of nodes, CyberBERT, a specialized Transformer variant trained on cybersecurity logs, is used to extract semantic representations of traffic data~\cite{govea2025hybrid}. This allows the system to capture long-range contextual dependencies within communication streams, such as subtle signals of a slow-rate DDoS attack or data-type probing~\cite{govea2025hybrid}. Note that GNNs can monitor the interaction graph while Transformers can analyze the content of communication, highlighting an intersection and an interesting opportunity to use AI to solve communication vulnerabilities.

AI is also a key technique in advancements in the mitigation of operational vulnerabilities. One of the most profound operational risks is the existence of unmanaged IoT devices. AI models have become foundational tools for device discovery, analyzing network traffic patterns to infer device types, roles, and functions without relying on static signatures. This enables near 100\% asset inventory accuracy, allowing CISOs to identify devices with Known Exploited Vulnerabilities (KEVs) and prioritize remediation based on business impact. Once discovered, AI can establish behavioral baselines for each device. Any deviation from the expected communication pattern, such as a factory sensor pinging a foreign server during off-hours, triggers an automated response. This transition from reactive to proactive security is essential for countering the speed of AI-powered attacks, such as the Aisuru botnet, which uses precision flooding to adapt its DDoS patterns in real time.

Regarding device vulnerabilities, the focus shifted toward TinyML and specialized lightweight deep learning architectures. The development of SecureNet-Lite, a compact CNN that undergoes post-training quantization to 8-bit integers, represents a significant milestone for device-layer security~\cite{ndlovu2025lightweight}. This process reduces the model's footprint while maintaining high detection accuracy. SecureNet-Lite demonstrated a 40\% faster inference time compared to a linear SVM, while achieving 92.1\% accuracy in detecting novel zero-day attack variants. This allows constrained devices to perform on-site threat detection without the latency or bandwidth costs of cloud offloading.

Spiking Neural Networks (SNNs), which mimic the event-driven processing of biological neurons, also offer a highly energy-efficient solution for real-time threat detection at the edge~\cite{nisha2025ai}. By only processing data when specific ``spikes'' or events occur, SNNs significantly reduce power consumption compared to traditional neural networks. This makes them ideal for battery-powered sensors that are susceptible to sleep-deprivation and resource-exhaustion attacks.

Mitigating device-class attacks is also a promising opportunity to apply AI. These attacks, like sleep deprivation, aim to exhaust batteries by keeping nodes in constant activity. Mitigation strategies can involve deploying lightweight anomaly detection models that monitor duty-cycle and power-consumption signatures. If a node exhibits abnormal wake-up patterns, the local AI can trigger a secure re-authentication or throttle requests to preserve the remaining energy budget. Similarly, resource exhaustion attacks targeting CPU and RAM are mitigated through Indicators of Attack (IoA) analysis, where AI identifies request patterns designed to overwhelm processing power.

Beyond the five vulnerability classes, two horizontal advancements have transformed IoT security. One important direction is the Explainable AI (XAI). As AI moves into critical decision-making roles, the ``black-box'' nature of deep learning becomes a liability~\cite{moss2025explainable}. XAI addresses this by making decisions transparent and traceable~\cite{moss2025explainable}. Using techniques such as SHAP and LIME, security platforms can provide feature importance charts that justify an alert~\cite{govea2025hybrid}. For instance, if a medical device is quarantined, the XAI module might explain the decision. This transparency is now a regulatory requirement in sectors like finance and healthcare, ensuring that human analysts can validate and audit AI-driven actions~\cite{moss2025explainable}.

A second important direction is Federated Learning (FL), which enables distributed security modeling without centralizing raw data. This is vital for privacy-sensitive applications such as healthcare and smart homes. In addition, this is well aligned with IoT requirements because devices and gateways often cannot afford to transmit sensitive logs continuously to a cloud server~\cite{devine2025federated,albanbay2025federated,shahin2025two}. Recent studies show that FL-based IDS can effectively detect IoT attacks while reducing privacy exposure and communication overhead, including for DDoS detection and privacy-preserving anomaly classification. Broader privacy surveys also identify federated learning, differential privacy, and homomorphic encryption as the dominant decentralized strategies for IoT protection~\cite{kaur2025survey}.

AI can evolve the IoT security from static, human-led defense to autonomous, AI-integrated resilience. Despite these advances, the use of AI in IoT security still faces major challenges. Notably, IoT data are highly non-IID and often subject to concept drift. This drift can degrade model accuracy over time. In addition, adversaries may exploit adversarial machine learning to evade or poison detection systems.

%% file: Conclusion.tex
\section{Conclusion}
\label{sec:Conclusion}

The exponential growth of the IoT applications has brought unprecedented technological advancements across critical sectors, but it has also drastically expanded the attack surface, a challenge exacerbated by device heterogeneity, lack of standardization, and inherent computational constraints. Recognizing that a fragmented approach to security is insufficient to protect these diverse scenarios, this survey presented a comprehensive, multi-dimensional analysis of the IoT threat landscape. Based on the literature, we detailed 28 common attacks, ranging from classic network interceptions to specialized physical exploits like node replication and skimming.

To bridge the gap between theoretical vulnerabilities and practical defense, this paper proposed a robust classification framework. Our application of the STRIDE model revealed that Information Disclosure and Denial of Service are the most prevalent functional threats, while the CVSS evaluation demonstrated a high-risk profile for IoT, with the vast majority of attacks classified as ``High'' or ``Critical'' severity. Furthermore, we introduced a novel mapping that correlates these 28 attacks with five foundational vulnerability classes (Process, Code, Communication, Operation, and Device) and positions them across the three core architectural layers of IoT applications. This multi-layered taxonomy provides stakeholders with an actionable tool for proactive defense planning and context-aware reactive monitoring, allowing for targeted mitigations ranging from hardware-based security modules to application-level access controls.

While current countermeasures, such as lightweight cryptography and intrusion detection systems, provide an essential baseline, the continuously evolving nature of cyber threats demands paradigm shifts. As highlighted in our discussion, the convergence of IoT with 5G networks and the increasing sophistication of botnets amplify the potential for large-scale disruptions. Consequently, future research must prioritize the integration of AI-assisted anomaly detection via Federated Learning, the widespread adoption of Zero Trust architectures, and the critical, yet challenging, migration to Post-Quantum Cryptography (PQC) in constrained environments.

Finally, this survey emphasized the potential of IoT over Named Data Networking (IoT-NDN) as a transformative, data-centric architecture that embeds security directly into the data packets, mitigating many inherent flaws of traditional IP-based communications. By consolidating threat identification, precise architectural mapping, and emerging technological trends, this survey serves as both a practical reference guide and a strategic roadmap for researchers and practitioners dedicated to engineering secure, resilient, and future-proof IoT ecosystems.

To build upon the findings of this survey, future research should prioritize addressing critical gaps in scalable threat detection and long-term device security. A primary direction can be the development of open-source, standardized frameworks for continuous IoT attack monitoring, which remains a significantly underexplored area. Furthermore, advancing AI-assisted anomaly detection through privacy-preserving approaches, such as Federated Learning, will be essential to counter large-scale botnets without compromising data confidentiality. As the computational landscape evolves, designing lightweight Post-Quantum Cryptography (PQC) and hybrid cryptographic schemes tailored for constrained environments is urgently needed to ensure long-term cryptographic agility. Finally, while IoT over Named Data Networking (IoT-NDN) offers a promising data-centric security paradigm, future studies must focus on resolving its specific open challenges, particularly mitigating interest flooding attacks, preventing content cache poisoning, and establishing scalable namespace-driven trust schemas for efficient key management.